\documentclass{ws}
\usepackage{dsfont,amsmath,bm,braket,color}
\usepackage{definitions}%

\begin{document}
\markboth{I.Fialkovsky, M.Zubkov}
	{Elastic Deformations And Wigner-Weyl Formalism In Graphene}
\title{ELASTIC DEFORMATIONS AND WIGNER-WEYL FORMALISM IN GRAPHENE}

\author{IGNAT V. FIALKOVSKY \footnote{On leave of absence from CMCC-Universidade  Federal  do  ABC,  Santo  Andre,  S.P.,  Brazil}}
\address{Physics Department, Ariel University, Ariel 40700, Israel\\
	ifialk@gmail.com}

\author{MIKHAIL A. ZUBKOV \footnote{On leave of absence from Institute for Theoretical and Experimental Physics, B. Cheremushkinskaya 25, Moscow, 117259, Russia}}
\address{Physics Department, Ariel University, Ariel 40700, Israel\\
		zubkov@itep.ru}

%\tableofcontents
\maketitle

\begin{history}
	\received{Day Month Year}
	\revised{Day Month Year}
	%\accepted{(Day Month Year)}
	%\comby{(xxxxxxxxxx)}
\end{history}

\begin{abstract}
	We discuss the tight-binding models of solid state physics with the $Z_2$ sublattice symmetry  in the presence of elastic deformations, and their important particular case -the tight binding model of graphene. In order to describe the dynamics of electronic quasiparticles we explore Wigner-Weyl formalism. It allows to calculate the two-point Green's  function in the presence of both slowly varying external electromagnetic fields and the inhomogeneous modification of the hopping parameters resulted from the elastic deformations. The developed formalism allows us to consider the influence of elastic deformations and the variations of magnetic field on the quantum Hall effect.
\end{abstract}
%\pacs{73.43.-f}
% system definitions

\section{Introduction}

Recently there has been the revival of interest to the Wigner-Weyl formalism in both condensed matter and high energy physics. It was proposed long time ago by H. Groenewold \cite{1} and J. Moyal \cite{2} mainly in the context of the one-particle quantum mechanics. The main notions of this formalism are those of the Weyl symbol of operator and the Wigner transformation of functions. Correspondingly, the formalism accumulated the ideas of H. Weyl \cite{3} and E. Wigner \cite{4}. In quantum mechanics the Wigner-Weyl formalism substitutes the notion of the wave function by the so called Wigner distribution that is the function of both coordinates and momenta. The operators of physical quantities are described by their Weyl symbols (that are also the complex-valued functions of momenta and coordinates). The product of operators on the language of Wigner-Weyl formalism becomes the Moyal product of their Weyl symbols \cite{5,berezin}.
The Wigner-Weyl formalism has been applied to several problems in quantum mechanics   \cite{6,7}). Notice that certain modifications of this formalism were proposed \cite{8,9,10,11,12,13,14,Buot}, where the main notions were changed somehow.

Let us recall the basic notions of the Wigner-Weyl formalism in quantum mechanics on the example of the one dimensional model. The Wigner distribution $W(x,p )$ is a function of  coordinate $x $ and momentum $p$. It gives the probability that the coordinate $ x $ belongs to the interval $[a,b]$ in the following way:
$$
{P} [a\leq x \leq b]=\frac{1}{2\pi}\int _{a}^{b}\int _{-\infty }^{\infty }W(x,p )\,dp\,dx
$$
Let $\hat A$ be operator of a certain physical observable. Its Weyl symbol $A_W(x, p)$ is defined as the function in phase space, which gives the expectation value of the given quantity with respect to the Wigner distribution $W(x,p )$ as follows \cite{2,15}
$$
{ \langle {\hat {A}}\rangle =\frac{1}{2\pi}\int A_W(x,p )W(x,p )\,dp\,dx.}
$$
For the pure quantum state the Wigner function is given by
$$
W(x,p ) = \int dy \, e^{-ipy} \psi^*(x+y/2) \psi(x-y/2)
$$
where $\psi(x)$ is the wave function of the state. The formalism is readily generalized to multidimensional  $\bx$ and $\bp$.

The Schrodinger equation in the language of the Wigner-Weyl formalism acquires the form $$
i \partial_t W(\bx,\bp ,t) = H_W(\bx,\bp )\ast W(\bx,\bp ,t)-W(\bx,\bp ,t)\ast H_W(\bx,\bp )
$$
The Moyal product of two functions $f$ and $g$ is given here by
$$
{f\ast g=f\,\exp {\left({\frac {i}{2}}({\overleftarrow {\partial }}_{\bx}{\overrightarrow {\partial }}_{\bp}-{\overleftarrow {\partial }}_{\bp}{\overrightarrow {\partial }}_{\bx})\right)}\,g}
$$
The left arrow above the derivative shows that the derivative acts on $f$ while the right arrow assumes the action of the derivative on $g$.

The definition of the Weyl symbol of  an arbitrary operator $\hat{A}$ is:
$$
A_W(\bx,\bp ) = \int \D\by e^{-i\bp\by} \langle \bx +\by/2| \hat{A} | \bx -\by/2\rangle
$$
We denote by $H_W(\bx,\bp )$ the Weyl symbol of Hamiltonian $\hat H$.

The Wigner-Weyl formalism was also modified somehow in order to be applied to the quantum field theory. The analogue of the Wigner distribution was introduced in QCD \cite{QCDW,QCDW2}. It has been used in the field-theoretic kinetic theory \cite{KTW,KTW2}, in noncommutative field theories \cite{NCW,NCW2}. Certain applications of the Wigner-Weyl formalism were proposed to several fields of theoretical physics including cosmology \cite{CSW,WW,Berry}.

The Wigner-Weyl formalism has been widely applied to the study of nondissipative transport phenomena \cite{ZW1,ZW2,ZW3,ZW4,ZW5,ZW6}. Using this formalism it has been shown that the response of nondissipative currents to the external field strength is expressed through the topological invariants that are robust to the smooth deformation of the system. This allows to calculate the nondissipative currents for certain complicated systems within the more simple ones connected to the original systems by a smooth deformation.
Using this method the absence of the equilibrium chiral magnetic effect \cite{CME} was demonstrated within the lattice regularized field theory \cite{ZW5}. The anomalous quantum Hall effect was studied using the Wigner-Weyl formalism for the Weyl semimetals and topological insulators \cite{ZW6}. In addition, the Wigner-Weyl formalism allows to  derive the chiral separation effect \cite{CSE} within the lattice models \cite{ZW3,ZW1}. The same method was also applied to the investigation of the hypothetical color-flavor locking phase in QCD \cite{ZW4}, where the fermion zero modes on vortices were discussed.   The scale magnetic effect \cite{SME} has also been investigated using the same technique \cite{ZW2}.

Historically the momentum space topological invariants were treated mainly in the context of condensed matter physics theory
\cite{HasanKane2010,Xiao-LiangQi2011,Volovik2011,Volovik2007,Volovik2010}.
They protect gapless fermions on the edges of the topological insulators \cite{Gurarie2011,EssinGurarie2011} and the gapless fermions in the bulk of Weyl semi-metals \cite{Volovik2003,VolovikSemimetal}. The fermion zero modes of various topological defects in $^3$He are also governed by momentum space topology \cite{Volovik2016}. In the high energy physics the topological invariants in momentum space were considered, say, in \cite{NielsenNinomiya1981,So1985,IshikawaMatsuyama1986,Kaplan1992,Golterman1993,Volovik2003,Horava2005,Creutz2008,Kaplan2011}.
The Wigner-Weyl formalism in \cite{ZW1,ZW3,ZW5,ZW6} was developed in the context of lattice field theory. The one-particle fermion Green's  function ${G}({\bp},\bq )$  was considered in momentum space $\cM$  (${\bp}, \bq  \in {\cM}$). It has been shown that the introduction of an Abelian external gauge field $\bA({x})$ resulting  in the Peierls substitution leads to the following equation
$$
{\hat Q}({\bp}-\bA(i {\partial_{\bp}}))){G}({\bp},\bq ) = \delta({\bp}-\bq )
$$
Here ${\hat Q}({\bp})$ is the lattice Dirac operator. Notice, that in lattice field theory the imaginary time is discretized on the same ground as space coordinates.

The Wigner transformation of Green's  function is defined as follows
\begin{equation}
\begin{aligned}
{G}_W({\bx},\bp)\equiv\int \D\bq  e^{i{\bx} \bq } {G}({{\bp}+\bq /2}, {{\bp}-\bq /2})\label{GWx0}
\end{aligned}
\end{equation}
It was shown \cite{ZW1} that for slowly varying external fields it obeys the Groenewold equation
\be
{G}_W({\bx},{\bp}) \ast Q_W({\bx},{\bp}) = 1 \label{Geq0}
\ee
with the above defined Moyal product $\ast$ extended to the $D$-dimensional vectors of coordinates ${\bx}$ and momentum $\bp$. Here $Q_W$ is the Weyl symbol of operator $ {\hat Q}({\bp}-\bA(i {\partial_{\bp}}))$.

In \cite{SZ2019,ZW2019,ZZ2019} the approach of
\cite{ZW1,ZW3,ZW5,ZW6} was further developed. In \cite{SZ2019} the lattice model with Wilson fermions was investigated in details. The precise expression for the Weyl symbol of the Wilson Dirac operator was derived in the presence of an arbitrarily varying external gauge field. In addition, the complete iterative solution of the Groenewold equation Eq. (\ref{Geq0}) was given. As a result the fermion propagator in the background of arbitrary external electromagnetic field was calculated. We refer to \cite{SZ2019} for the technical details of the Wigner-Weyl formalism in lattice models, which will be used in the present paper as well.
In \cite{ZW2019} it has been shown that in the lattice models (i.e. in the tight-binding models) of solid state physics with essential inhomogeneity (caused by the varying external magnetic field) the Hall conductance integrated over the whole space is given by a topological invariant in phase space. This quantity is expressed through the Wigner transformation of the one-fermion  Green's  functions. The expression for the phase space topological invariant repeats the form of the momentum space topological invariants of \cite{Matsuyama:1986us,Volovik0,Gurarie2011,EssinGurarie2011,ZW1}. The difference is that the Green's  functions entering this expression depend on both momenta and coordinates, and the ordinary product is substituted by the Moyal product, while the extra integration over the whole space is added \footnote{The topological invariants of \cite{Matsuyama:1986us,Volovik0,Volovik2003,Gurarie2011,EssinGurarie2011} repreat the structure of the degree of mapping of the three-dimensional manyfold to a group of matrices.}. It has been shown that the value of the  topological invariant in phase space responsible for the Hall conductance is robust to the introduction of disorder. Certain indications were found that it is also robust to the weak Coulomb interactions.

Topological description of the Quantum Hall effect (QHE) started from the discovery of the TKNN invariant  \cite{TTKN} defined in the two-dimensional systems. The three dimensional topological invariants for the QHE were considered in \cite{Hall3DTI}. This formalism allows to deal with the intrinsic anomalous quantum Hall effect (AQHE) and with the QHE in the presence of {\it constant} magnetic field \cite{Fradkin}. Unfortunately, the formalism that is based on the notion of Berry curvature does not admit the direct generalization to the QHE in the presence of varying external magnetic field and elastic deformations when the system becomes essentially inhomogeneous. It is widely believed that the total QHE conductance is robust to the introduction of disorder and weak interactions.
Expression for the QHE conductivity through the one-particle Green's  functions has been invented in \cite{Matsuyama:1986us,Volovik0}. In the presence of interactions the full two-point Green's  function should be substituted to the corresponding expression. It has been checked in \cite{ZZ2019}, that the leading order contributions due to the Coulomb interactions do not change expression for the AQHE conductivity in topological insulators. This expression has the form of an integral in momentum space over the certain composition of the interacting two-point Green's  function. It is worth mentioning, however, that there is still no proof in general case to all orders in perturbation theory that the higher order full Green's  functions do not give contributions to the QHE.  For a discussion of this issue see also \cite{Gurarie2011,EssinGurarie2011}. The AQHE conductivity discussed in \cite{ZZ2019} may be applied, in particular, to Weyl semimetals \cite{semimetal_effects10,semimetal_effects11,semimetal_effects12,semimetal_effects13,Zyuzin:2012tv,tewary}.

It is difficult to overestimate the role of disorder in the Quantum Hall Effect \cite{Fradkin,TKNN2,QHEr,Tong:2016kpv}. One of its effects is elimination of the Hall current from the bulk, and its concentration along the boundary. The formalism developed in \cite{ZW2019} allows to give an alternative prove that the total conductance remains robust to the introduction of disorder in the majority of systems although the total current remains only along the boundary of the sample. However, for graphene there are certain complications to be discussed in Conclusions. Namely, when the Hall current remains along the boundary only, the QHE is absent at the half filling (neutrality point).
 According to the common lore the Hall conductance is assumed to be robust to the introduction of weak interactions, at least in the presence of the sufficient amount of disorder. However, Coulomb interactions are able to give rise to the fractional QHE \cite{Fradkin,Tong:2016kpv,Hatsugai} for the clean systems at very small temperatures. This, however, is out of the scope of the present paper.

Graphene \cite{vozmediano2,vozmediano4,vozmediano5,vozmediano6,Oliva,VZ2013,VolovikZubkov2014,Volovik:2014kja,Khaidukov:2016yfi}
represents the two-dimensional Weyl semimetal. The low energy physics of its electronic quasiparticles is described by massless Dirac equation. Therefore, it allows to simulate in laboratory certain features of the high energy physics that cannot be observed directly. The examples of such effects are: the Schwinger pair production, and the gravitational effects in the quantum-mechanical motion of particles. Gravity appears in graphene in the presence of elastic deformations \cite{VZ2013}. One more exceptional feature of graphene is that (unlike discovered later three-dimensional Weyl semimetals) it is described with a very good accuracy by the simple tight-binding model defined on the honeycomb lattice. The investigation of various features of this model (including the QHE) based on the Wigner-Weyl formalism constitutes the subject of the present paper.  It worth noting here, that many phenomena in graphene can be adequately treated even with low energy continuum approximation,  within appropriate (pseudo-relativistic) field-theoretical methods \cite{Fialkovsky:2016kio,Fialkovsky:2011wh}.

In graphene the elastic
deformations lead both to the appearance of the emergent gauge field and emergent
gravitational field (see, for example, \cite%
{VZ2013,Horava2005,VZ2014NPB,WeylTightBinding,Cortijo:2015jja,Chernodub:2015wxa,Zubkov:2015cba}%
).  The emergent gauge field appears as the variation of the
Fermi point position in momentum space while the emergent gravitational field comes as  the variation of
the slope of the dependence of energy on momentum (i.e. the anisotropic Fermi velocity).

Although  our main aim is the investigation of the tight-binding model of monolayer graphene, the paper is organized in such a way, that many of the obtained expressions may be applied to some other lattice models of solid state physics (though, only the model of graphene from this class describes quantitatively the really existing physical system).
The paper is organized as follows. We start from the description of the almost arbitrary non-homogeneous lattice model in Sect. \ref{Nonlocal}. We represent the formulation of such models in momentum space. Next, we reduce the considered class to the tight-binding models with the jumps of electrons between the adjacent lattice sites only.  This section is ended with the consideration of non-homogeneous tight-binding models with the $Z_2$ sublattice symmetry. Tight-binding model of graphene belongs to this class. However, it is much wider, in particular the tight-binding models defined on rectangular lattices in $2D$ and $3D$ remain in this class.

In Sect. \ref{WignerWeylLattice} we introduce the Wigner-Weyl formalism in the nonhomogeneous lattice models with $Z_2$ sublattice symmetry. We explore the  definition of the Weyl symbol of lattice Dirac operator (entering the fermion action), which is defined through the integral in momentum space. We calculate the Weyl symbol of Dirac operator for the considered models both in the presence of inhomogeneous hopping parameters and in the presence of varying external electromagnetic field. Both electromagnetic field and the hopping parameters are assumed to vary slowly, i.e. we neglect their variations on the distance of the lattice spacing.
Next, we  turn directly to the physics of graphene. We recall the relation between elastic deformations and the non-homogeneous hopping parameters. After that we express the Weyl symbol of lattice Dirac operator in graphene in the presence of elastic deformations and electromagnetic field through the electromagnetic potential and the tensor of elastic deformations.

In Sect. \ref{SectWW} we consider relation between Weyl symbol of lattice Dirac operator in the considered systems and the Wigner transformation of the Green's  functions. Also we express electric current through the quantities of the Wigner -Weyl formalism.

In Sect. \ref{GWcalculation} we again consider the lattice models of general type. (The corresponding calculations are of course applied to the case of graphene directly.) Namely, we extend the results of \cite{SZ2019} for the calculation of the Wigner transformation of the fermion Green's  function (obtained for lattice Wilson fermions on rectangular lattice) to the case of the non-homogeneous tight-binding models of arbitrary form. It is explained also how to reconstruct the Green's  function both in momentum and coordinate representations from its known Weyl symbol.

In Sect. \ref{sigmaTop} we extend the consideration of \cite{ZW2019} to the case, when elastic deformations are present. Namely, we prove that for the {\it noninteracting} $2D$ condensed matter model with {\it slowly varying electromagnetic fields and elastic deformations} the Hall conductivity integrated over the whole area of the sample is given by topological invariant in phase space composed of the Wigner transformed one-particle Green's  function. It is the same topological invariant proposed in \cite{ZW2019}. It remains robust to the smooth modification of the model (if the modification remains local and bounded to the smooth modification of the Hamiltonian in the limited region of the sample that remains far from its boundary).

In Sect. \ref{IQHE} we apply the results of the previous sections to the discussion of Hall conductivity in graphene in the presence of both elastic deformations and inhomogeneous magnetic field. First, we recall the standard derivation of the Hall conductance in the noninteracting $2D$ models with constant magnetic field and constant hopping parameters. Next, this standard derivation is extended to the case of the weakly varying elastic deformations that cause varying hopping parameters that remain isotropic (i.e. their values are equal for all directions in the given point though vary from point to point). We obtain the formula for the Hall conductance that allows to express it through the total number of electrons in the occupied energy levels and the external magnetic field. Next, we apply the topological invariant in phase space defined in Sect. \ref{WignerWeylLattice} to the consideration of the QHE in graphene. The very existence of such a representation for the QHE conductance allows to prove that it remains robust to the weak elastic deformations of arbitrary form and weak modification of magnetic field unless the topological phase transition is encountered. Both are assumed to be localized in the region that remains far from the boundaries of the sample. Finally, in this section we notice that the elastic deformations in graphene that do not cause emergent magnetic field give rise to the isotropic hopping parameters. The corresponding displacement appears to be analytical function of the atom coordinates of the unperturbed honeycomb lattice. For the constant external magnetic field this allows to derive the simple relation between the number of electrons in the occupied branches of spectrum, and the value of magnetic field.

In Sect. \ref{SectConcl} we end with the conclusions, discuss the obtained results and the directions of future research.

Throughout the paper the following notational conventions are used.  Latin letters in subscript $a,b,c$ numerate the spatial components of vectors. The Latin letters in superscript $i,j$ enumerate the elementary translations. All momenta vectors are bold italic $\bl,\bk,\bp,\bq$ from the middle of the alphabet, coordinate vectors are from its end, $\bx,\by ,\bu,\bv$.
Operators are denoted by the Latin letters with hat $\hat Q, \hat G$, their matrix elements of operators - by functions of two variables $Q(\bp,\bq)$. Weyl symbols of operators are denoted by the sub-index $W$:
$ (\hat Q)_W \equiv Q_W $.

\section{Hamiltonian for the nonlocal tight--binding model}
\label{Nonlocal}
\subsection{General case}

We start our discussion with the general case of the non-local tight--binding model in presence of external electromagnetic field $A$. The discussion of the present section is applicable, in principle, not only to the tight-binding model of graphene, but also to other $2+1$D and $3+1$D tight-binding models of solid state physics.

The Hamiltonian under consideration has the form
\be
\begin{split}
\cH
& \equiv\sum_{\bx ,\by}\,\, \bar \Psi(\by) f(\by,\bx ) e^{\ii  \int_\bx ^\by d\bv  \bA( \bv)} \Psi(\bx )
\\
& = \frac{1}{|{\cM}|^2}\int d\bp  d\bq  d \bl  d \bk \,\, \bar \Psi(\bp )f(\bq , \bl ) \Psi ( \bk )
\sum_{\bx , \by}\, e^{\ii  \int_\bx ^\by d \bv  \bA( \bv)} e^{\ii \by(\bq -\bp )+i\bx  (- \bl + \bk )}
\label{bPsi-Psi-gen}
\end{split}
\ee
here the sum is over the lattice sites $\bx, \by$, while $f(\bx, \by)$ is the matrix of hopping parameters. The lattice is assumed to be infinite, which means that we neglect the finite volume effects as well as the finite temperature corrections.  Thus,  the integrals in the second line are over momentum space $\cM$, which is the first Brillouin zone specific for the given lattice model.

In the following we may absorb the electromagnetic field to the definition of $f$. Therefore, we omit it temporarily and will restorer in appropriate expressions.

%\subsection{Hamiltonian and operator $\hat Q$ in momentum representation}

Now let us consider a less general situation of the tight-binding model with the jumps of electrons between the adjacent sites only -- the nearest neighbor approximation. We discuss the case of the inhomogeneous hopping parameters, which will allow us to discuss elastic deformations.
Now
\begin{equation}
f(\by, \bx) = \sum_{j=1}^M \delta(\by- (\bx+\bbj)) f^{(j)}(\by) \label{fg}
\end{equation}
where $\bbj$ are the vectors connecting each atom to its nearest $M$ neighbors, $j = 1,...,M$,  and $f^{(j)}(\by) $ is the non-uniform varying hopping parameter.

Then
\be
\begin{split}
f(\bq , \bl)
& = \frac{1}{|{\cM}|}\sum_{j=1}^M \sum_{\bx , \by } e^{-\ii \bq \bx  + \ii  \bl  \by }
	\delta(\by-(\bx+\bbj ))f^{(j)}(\by) \\
&= \frac{1}{|{\cM}|}
		\sum_{j=1}^M \sum_\by e^{-\ii (\bq- \bl) \by  +\ii \bq  \bbj }f^{(j)}(\by)\\
&= \sum_{j=1}^M  {f}^{(j)}(\bq- \bl) e^{ \ii \bq  \bbj }
\end{split}
\label{f-fourier}
\ee
and %{\bf check sign of $ l$!!}
\be
\begin{split}
\cH &= \frac{1}{|{\cM}|}
 \sum_{j=1}^M \int_\cM \D{\bp  d\bq}   \bar \Psi(\bp)
\[ {f}^{(j)}(\bp-\bq) e^{ \ii \bq  \bbj }\]
		\Psi(\bq) 		
\end{split}
\ee

\subsection{The $Z_2$ sublattice symmetry}

Our next simplification is consideration of a particular case, when crystal lattice exhibits $Z_2$ sublattice symmetry, i.e. there are two sublattices $\cO_{1,2}$ that constitute the crystal, and there is the one to one correspondence between them generated by  shift $\bx \to \bx + \bbj$ for any $j = 1,...,M$ and $\bx \in \cO_1$ or $\bx \to \bx-\bbj$ for $\bx \in \cO_2$.% We took $ \bx =(t,\bx)$, $\bbj =(0,\bbj)$.

The points of those two sublattices are to be considered independently, which gives the sublattice index $\alpha=1,2$ to $\Psi$.   We identify $\Psi(t,\bx)$, where $\bx \in  \cO_1$, with $\Psi_1(t, \bx)$, and $\Psi(t,\bx)$ for $\bx \in \cO_2$ is identified with $\Psi_2(t,\bx)$. We set $\Psi_1(t,\bx) = 0$ for $\bx \in \cO_2$ and $\Psi_2(t,\bx) = 0$ for $\bx \in \cO_1$.  Then
$$
\Psi_\al(\bp) = \frac{1}{|{\cM}|^{1/2}} \sum_{\bx \in \cO_\al} \Psi_\al(\bx)e^{-\ii \bp \bx },
	\qquad
	\al=1,2
$$
The inverse transform is similarly
$$
	\Psi_\al(\bx) =\frac{1}{|{\cM}|^{1/2}} \int_\cM \D{\bp} \Psi_\al(\bp)e^{\ii  \bp \bx }, \qquad
	\al=1,2
$$
Note, that Brillouin zone  $\cM$ is the same for both sublattices since both of them are build over the same basis vectors, and thus have the same periodicity. $\cM$ is formed as space of vectors $\bp$ defined modulo transformations
$$
	\bp \to \bp + { \bg}^{(k)}
$$
where ${ \bg}^{(k)}$ are vectors of inverse lattice that solve the system of equations
\begin{eqnarray}
	e^{\ii { \bg}^{(k)}{\bmm}^{(j_1j_2)}}=1,\quad j_1,j_2 =  1, ..., M
\end{eqnarray}
while ${\bmm}^{(j_1,j_2)} =  \bb^{(j_1)}- \bb^{(j_2)}$ form each of the two sublattices $\cO_{1,2}$.
Then the hoping parameters $f^{(j)}$ become  $2\times 2$ matrices. Besides, we assume, that the spatial hoping parameters are coordinate dependent,
\be
f^{(j)}_{21}(\bx + \bbj )   = -t^{(j)}(\bx+ \bbj)
\ee
i.e., the values of $t^{(j)}$ may vary independently but not with time. The diagonal ones are  vanishing, $f^{(j)}_{11} = f^{(j)}_{22} = 0$.

The Hamiltonian then receives the form:
\begin{eqnarray}
\cH &=&
%\sum_{\alpha=1,2}\sum_{\bx \in \cO_1\cup \cO_2}
%	\frac{1}{2\Delta \tau} 	
%	\Big(
%		\bar{\Psi}_\alpha(t+\Delta \tau,\bx)\Psi_\alpha(t,\bx)-\bar{\Psi}_\alpha(t,\bx)\Psi_\alpha(t+\Delta \tau,\bx)
%	\Big)  \nonumber\\
%&=&
	\sum_{j=1}^M\sum_{\bx \in \cO_1\atop \by = \bx + \bbj}
	\Big(
		t^{(j)}(\by)\bar{\Psi}_2(t,\by) \Psi_1(t,\bx) + t^{(j)}(\by)\bar{\Psi}_1(t,\bx) \Psi_2(t,\by)
	\Big)
	\label{SLAT}
\end{eqnarray}
In what follows we will omit the temporal argument of the wave function whenever no confusion is provoked.
%\red{What if we define $\Psi_2(\bx)\equiv \Pis(x+\bbj )}

\section{Weyl symbol for the lattice Dirac operator}
\label{WignerWeylLattice}

\subsection{Lattice Dirac operator}

Let us rewrite (\ref{SLAT}) in the following way
\be
\begin{split}
\cH&=\sum_{{\bx \in \cO_1}\atop{\by \in \cO_2}}
		\(\bar{\Psi}_1(\bx),\ \bar\Psi_2(\by) \)
		 {\bm H}(\bx,\by )
		\({\Psi}_1(\bx),\ \Psi_2(\by) \)^T
	\\
	&\equiv
	\sum_{{\bx \in \cO_1}\atop{\by \in \cO_2}}
	\Big(
		\bar{\Psi}_2(\by) H_{21}(\by,\bx ) \Psi_1(\bx)+
		\bar{\Psi}_1(\bx) H_{12}(\bx,\by ) \Psi_2(\by)
	\Big)
\end{split}
\label{cH}
\ee
where
\be
\begin{split}
H_{21}(\by_2,\by_1)
& = -\sum_{j=1}^M \delta\(\by_2-(\by_1+\bbj )\)
	\tj \(\frac{\by_1+\by_2}{2}\),\qquad
	{{\by_1 \in \cO_1}\atop{\by_2 \in \cO_2}}\\
H_{12}(\by_1,\by_2)
& =  H_{21}(\by_2,\by_1)
\end{split}
\label{H21 t}
\ee
Note that we define the hoping parameter by its values in the middle of the lattice links, $\tj \(\tfrac{\by_1+\by_2}{2}\)$ for a better readability of consequent formulas.
	
We will refer to $2$ by $2$ matrix operator  ${{\bm H}}$ as to the lattice Dirac Hamiltonian, although its geometrical symmetries will only be defined after specifying $\bbj$. Along with Hamiltonian, we also introduce Dirac operator, which enters the action and consequently will be usefull for analysis of the partition function of the system
\be
	Q\equiv \ii \omega -{\bm H} =
	\(\begin{array}{cc}
	\ii \omega &-H_{12}\\
	- H_{21} &		\ii \omega\\
	\end{array}\).
\label{def Q}
\ee

%\subsection{The second definition of Weyl symbol-through the integral in momentum space}

Let us consider the off-diagonal term $21$ in the Hamiltonian. It can be written in terms of the Fourier transformation as
\be
\cH_{21}=\frac{1}{|{\cM}|}\int_\cM \D{\bp  d\bq }
\bar{\Psi}_2(\bp) H_{21}(\bp,\bq) \Psi_1(\bq)
\ee
modifying Eq. (\ref{f-fourier}) for two sublattices, we have
	\be
	\begin{split}
		H_{1 2}(\bq, \bp)
		& = \frac{1}{|{\cM}|}\sum_{j=1}^M \sum_{\by_1\in\cO_1\atop \by_2\in\cO_2} e^{-\ii \bq \by _1 + \ii \bp  \by _2}
		\delta\(\by_2-(\by_1+\bbj )\)
		\tj \(\frac{\by_1+\by_2}{2}\)\\
		&= \frac{1}{|{\cM}|}
		\sum_{j=1}^M \sum_{\by_1\in\cO_1} e^{-\ii (\bq -\bp ) \by _1 +\ii \bq  \bbj }
		\tj \(\by_1+\bbj /{2}\)
	\end{split}
	\label{f-fourier2}
	\ee
We can identify now the points $\by_1+\bbj /{2}$, $\by_1\in\cO_1$ with those situated in the middle of the lattice links along the $j$-th direction. We call these sets of points by $\cO^{(j)}_{{1}/{2}}$.
Now we can write
\be
	H_{21}(\bp,\bq) =  \sum_{j=1}^M \tj (\bp-\bq) e^{   \ii (\bp + \bq) \bbj /2}
\label{H_12}
\ee
where
\be
\tj(\bp) = \frac{1}{{|{\cM}|}} \sum_{\bx \in \cO^{(j)}_{{1}/{2}}} \tj(\bx) e^{-\ii \bx \bp }
\label{tj}
\ee
Note, that the above expression is simply a Fourier transformation of a function shifted by $\bbj /2$ since $\tj$ are only defined in the middle of the links.
In the particular case, when $t^{(j)}(\bx) = t^{(j)}$, i.e. if it does not depend on $ \bx $, we obtain
$$
	{t}^{(j)}(\bp) =t^{(j)} \delta(\bp\!\!\!\mod \bgj).
$$

\subsection{The definition of the Weyl symbol in momentum space}
We propose the following definition of the Weyl symbol of an operator $\hat A$:
\be
(\hat A)_{W} (\bx,\bp )
	 = \int_{{\cM}} \D\bq A(\bp+\bq /2,\bp-\bq /2)  e^{\ii  \bq  \bx }.
	 \label{A_W}
\ee
Here integral is over momentum space ${\cM}$, in which the two vectors are equivalent if they differ by $\bgj$. In particular this means that $\bp\pm \bq/2$ do not span the whole Brillouin zone $\cM$.

For off-diagonal components of ${\bm H}$ from above it gives
\be
H_{21,W} (\bx,\bp )
	= \int_{{\cM}} \D\bq  e^{\ii  \bq  \bx }\sum_{j=1}^M   \tj (\bq)\,  e^{  \ii \bp   \bbj  }
	= \sum_{j=1}^M  e^{\ii   \bp \bbj }   \int_{{\cM}} \D\bq   \tj (\bq) e^{\ii  \bq  \bx }
\ee
If the hopping parameters are homogeneous, then
$$
H_{21,W} (\bx,\bp ) =  e^{\ii   \bp \bbj }  \tj
$$	
On the other hand, when the hopping parameters vary, we have using \Ref{tj}
\be
\begin{split}
H_{21,W} (\bx,\bp )
%		&= \sum_j  e^{\ii   p \bbj }   \int_{{\cM}} d\bq  \,  \tj  (\bq) e^{\ii  q  \bx   }\\
	&=  \frac{1}{|{\cM}|}   \sum_{j=1}^M  e^{\ii   \bp \bbj }  \sum_{\by \in O^{(j)}_{1/2}} \tj(\by)
		\int_{{\cM}} \D\bq  \,    e^{\ii  \bq  (\bx-\by)  }\\
&=  \sum_{j=1}^M  e^{\ii   \bp \bbj }  \sum_{\by \in \cO^{(j)}_{1/2}} \tj(\by) {\cal F}(\bx-\by ),
\end{split}
\label{H21W}
\ee
where
\be
	{\cal F}(\bx)  = \frac{1}{|{\cM}|} \int_{{\cM}} \D\bq  e^{\ii  \bq  \bx  }
\ee
Notice that for $\bx, \by \in \cO^{(j)}_{1/2}$ we have $\by-\bx\in\cO$ and thus, the function ${\cal F}(\by-\bx)$ vanishes for all $ \bx \in \cO^{(j)}_{1/2}$ except for $ \bx =\by$. However, it remains nonzero and oscillates for all other values of $ \bx $, including continuous ones, and gives unity if summed over $\cO^{(j)}_{1/2}$ for any $ \bx $
\be
	\sum_{\by \in \cO^{(j)}_{1/2}} {\cal F}(\bx-\by ) = 1.
\ee
Each term of the $j$-sum in \Ref{H21W} receives a particular form if $ \bx  \in \cO^{(j)}_{1/2}$ (with same value of $j$):
\be
	H^{(j)}_{21,W} (\bx,\bp )\Big|_{\bx\in \cO^{(j)}_{1/2}} = 	e^{\ii \bp \bbj }  \tj(\bx )
\ee
However, \Ref{H21W} defines $H_W$ also for the continuous values of $ \bx $.

The presence of external electromagnetic field with vector potential $ \bA$ may be introduced to the model via the modification of hopping parameter in the term $H_{21}$
$$
	\tj(\bx) \to \tj(\bx) e^{-\ii \int_{\bx-\bbj /2}^{\bx + \bbj /2}  \bA(\by)d\by }.
$$
In $H_{12}$ there should be a complex conjugate substitution:
 $$
\tj(\bx) \to \tj(\bx) e^{-\ii \int_{\bx + \bbj /2}^{\bx-\bbj /2}  \bA(\by)d\by }
$$
From \Ref{H21W} we see that it is simply Pieirls substitution in the language of Weyl symbols.

Combining this substitution with \Ref{def Q} we get
$$
	Q_W = \sum_{j=1}^{M} Q^{(j)}_W
$$
where
\begin{equation}
Q^{(j)}_W(\bx,\bp )\Big|_{\bx \in  \cO^{(j)}_{1/2} }
	= \left(\begin{array}{cc}
		\ii \omega/M & -\tj (\bx)\, e^{\ii  (\bp  \bbj-{A}^{(j)}(\bx))} \\
		-\tj(\bx) \, e^{-i (\bp \bbj-{A}^{(j)}({\bx} ))} & \ii \omega/M
	\end{array}\right)
\label{QWj},
\end{equation}
 $M$ is the number of the nearest neighbours. Here
$$
	A^{(j)}(\bx) = \int_{\bx-\bbj /2}^{\bx + \bbj /2}  \bA(\by)d\by.
$$
For both $\tj$ and $ A$ that almost do not vary at the distances of order of lattice spacing we may use Eq. \Ref{QWj} for arbitrary values of $ \bx $, and get
\begin{equation}
Q_W(\bx,\bp )=\sum_{j=1}^{M}
	\left(\begin{array}{cc}
		\ii \omega/N & -t^{(j)}(\bx)\, e^{\ii  (\bp  \bbj-{A}^{(j)}(\bx))} \\
		-t^{(j)}(\bx) \, e^{-\ii (\bp \bbj-{A}^{(j)}(\bx))} & \ii \omega/N
	\end{array}\right).
 \label{QW}
\end{equation}
This approximation corresponds to the situation, when the typical wavelength of electromagnetic field is much larger than the lattice spacing.

%\section{Elastic deformations and Weyl symbol of lattice Dirac operator}
%\label{Graphene}

\subsection{Elastic deformation and modification of hoping parameters}

Now we are in the position to consider elastic deformations and Wigner-Weyl formalism in graphene. In this section we discuss graphene monolayer in the presence of elastic deformations. The sheet of graphene is parametrized by coordinates $ x_k, k = 1,2$. The displacements of each point have three components $u_a(\bx)$, where $a=1,2,3$. The resulting coordinates of the graphene sheet embedded into three-dimensional space $y_a$ are given by
\begin{eqnarray}
y_k(\bx)& = & x _k + u_k(\bx), \quad k = 1,2\nonumber\\
y_3(\bx)& = &u_3(\bx)
\end{eqnarray}
In the absence of the displacements, when $u_a=0$, the graphene is flat. Metric of elasticity theory is given by
\begin{equation}
g_{ik} = \delta_{ik} + 2 u_{ik},~~u_{ik} = \frac{1}{2}\Bigl(\partial_i u_k +
\partial_k u_i +  \partial_i u_a \partial_k u_a\Bigr), \quad a = 1,2,3,
\quad i,k = 1,2.
\label{Two_deformations}
\end{equation}
Elastic deformations change the spatial hopping parameters which enter \Ref{H21 t} are now
% \be
% Q_{21}(\by_2,\by_1) = \sum_j \delta\(\by_2-(\by_1+\bbj )\)
% 	\tj \Big((\by_2+y_1)/2\Big),\qquad
% 	{{\by_1 \in \cO_1}\atop{y_2 \in \cO_-}}
%\ee
%where
\begin{equation}
t^{(j)}(\bx)=t\(1-{\beta} u_{ik}(\bx)  b_i^{(j)} b_k^{(j)}\).
\label{HoppingElements}
\end{equation}
Here
\be
\{\bbj \}_{j=1}^3 = \{(-1,0); (1/2,\sqrt{3}/2); (1/2,-\sqrt{3}/2)\}
\label{l-s-gr}
\ee
 while $\beta$ is the Gruneisen parameter. We imply that $\beta |u_{ij}| \ll 1$.

The standard expression for the emergent electromagnetic potential has the form
\begin{eqnarray}
{A}_1  & = &- \frac{\beta}{a}\,u_{12}\nonumber\\
{A}_2  & = & \frac{\beta}{2a}\,(u_{22}-u_{11})\label{AFP2_}
\end{eqnarray}

For arbitrarily varying field $u$ we obtain the following expression for $Q_W$:
\be
Q_W(\omega,p; \tau, \bx ) =
\ii \omega -t  \sum_{j=1}^3
	\( 1- \beta u_{ik}(\bx)  b_i^{(j)} b_k^{(j)} \)
	\(\begin{array}{cc}
		0 & e^{\ii  (\bp  \bbj -{ A}^{(j)}(\bx))} \\
		e^{-\ii (\bp  \bbj -{ A}^{(j)}(\bx ))} & 0
	\end{array}	\)\label{QW}
\ee

\section{Green's  function and the Groenewold equation  }
\label{SectWW}
\subsection{Appearance of the Moyal product }

\label{sec:double-lattice}

Our definition of the Weyl symbol of operator $\hat A$ \Ref{A_W} can be also written as
\begin{eqnarray}
A_W(\bx,\bp ) &=&
\int_{{\cM}}d\bcP e^{ix\bcP}
\Bra{\bp+\frac\bcP{2}} \hat{A}  \Ket{\bp-\frac\bcP{2}}
\end{eqnarray}
The integral over $\bcP$ is over the Brillouin zone ${\cM}$, i.e. in ${\cM}$ we identify the points that differ by  a vector of reciprocal lattice $\bgj$.

%In the case of the tight-binding models considered in the present paper matrix elements of the considered operators   $\Bra{p+\bcP/2} \hat{A}  \Ket{p-\bcP/2}$ depend on
%$p \,{\rm mod}\, {\bf G}_j$ and $\bcP\, {\rm mod}\, {\bf G}_j$. Therefore,
%\begin{eqnarray}
%A_W(\bx,\bp ) &=&
% \int_{{\cM}}d\bcP e^{ix\bcP}
%\Bra{p+\frac\bcP{2}} \hat{A}  \Ket{p-\frac\bcP{2}}
%\end{eqnarray}
%according to the above given definitions.

Now let us consider the Weyl symbol $(AB)_W(\bx,\bp )$ of the product of two operators $\hat A$ and $\hat B$ such that their matrix elements $\Bra{\bp+\frac\bq {2}} \hat{A} \Ket{\bp-\frac\bq {2}}$ and $\Bra{\bp+\frac\bq {2}} \hat{B} \Ket{\bp-\frac\bq {2}}$ are nonzero only when $\bq$ remains in the small vicinity of zero. Then
\begin{equation}\begin{aligned}
&(AB)_W(\bx,\bp )=
 \int_{{\cM}} \D{\bcP} \int_{\cM} \D{\bcR}
 	e^{\ii \bx \bcP}
 		\Bra{\bp+\tfrac\bcP{2}} \hat{A} \Ket{\bcR}
		\Bra{\bcR} \hat{B} \Ket{\bp-\tfrac\bcP{2}}\\
&=\frac{1}{2^D}\int_{{\cM}} \D{\bcP d\bcK }
	e^{\ii \bx \bcP}
		\Bra{\bp+\tfrac\bcP{2}} \hat{A} \Ket{\bp-\tfrac\bcK{2}}
		\Bra{\bp-\tfrac\bcK{2}}\hat{B} \Ket{\bp-\tfrac\bcP{2}}\\
&= \frac{2^D}{2^D }\int_{{\cM}} \D{\bq d \bk} e^{\ii \bx (\bq+ \bk)}
	\Bra{\bp+\tfrac\bq {2}+\tfrac{ \bk}{2}} \hat{A} \Ket{\bp-\tfrac\bq {2}+\tfrac{ \bk}{2}}
	\Bra{\bp-\tfrac\bq {2}+\tfrac{ \bk}{2}}\hat{B} \Ket{\bp-\tfrac\bq {2}-\tfrac{\bk}{2}}\\
&= \int_{{\cM}} \D{\bq d \bk}
	\[  e^{\ii \bx  \bq}
		\Bra{\bp+\tfrac\bq {2}} \hat{A} \Ket{\bp-\tfrac\bq {2}}
	\]
	e^{\tfrac{ \bk}{2}\cev{\partial}_\bp-\tfrac\bq {2}\vec{\partial}_\bp}
	\[  e^{\ii \bx   \bk}
		\Bra{p+\tfrac{ \bk}{2}}\hat{B} \Ket{\bp-\tfrac{ \bk}{2}}
	\]\\
&= \[ \int_{{\cM}} \D\bq  e^{\ii \bx  \bq}
		\Bra{\bp+\tfrac\bq {2}} \hat{A} \Ket{\bp-\tfrac\bq {2}}
	\]
	e^{\tfrac{\ii}{2} \(- \cev{\partial}_\bp\vec{\partial}_\bx+\cev{\partial}_{\bx}\vec{\partial}_\bp\)}
	\[ \int_{{\cM}} \D\bk e^{\ii \bx   \bk}
		\Bra{\bp+\tfrac{ \bk}{2}}\hat{B} \Ket{\bp-\tfrac{ \bk}{2}}
	\]
\label{Z}
\end{aligned}
\end{equation}
Here the bra- and ket- vectors in momentum space are defined modulo vectors of reciprocal lattice $\bgj$,  as it is inflicted by the periodicity of the lattice. In the second line we change variables
$$
	\bcP = \bq+ \bk , \quad \bcK = \bq- \bk
$$
$$
	\bq = \frac{\bcP+\bcK}{2}, \quad  \bk =\frac{\bcP-\bcK}2
$$
with the Jacobian
$$
J = \left|\begin{array}{cc} 1 & 1 \\
-1 & 1 \end{array} \right| = 2^D
$$
This results in the factor ${2^{D}}$ in the third line. Here $D$ is the dimension of space. In the present paper it may be either $2$ or $3$.

Hence, the  Moyal product  may be defined similar to the case of continuous space
\begin{equation}\begin{aligned}
&(AB)_W(\bx,\bp )=
 A_W(\bx,\bp )
	e^{\frac{\ii}{2} \( \cev{\partial}_{\bx}\vec{\partial}_\bp-\cev{\partial}_\bp\vec{\partial}_{\bx}\)}
B_W(\bx,\bp )
\label{ZAB}\end{aligned}\end{equation}
Notice, that for the chosen form of Wigner transformation on a lattice the above equality is approximate and works only if the operators $\hat{A}$, $\hat{B}$ are close to  diagonal.

\subsection{Lattice Groenewold equation}
\label{Sec:GroEq}

Let us define the Fourier components of field $\Psi(\tau,\bx)$ that depends on both space coordinates $\bx$ and imaginary time $\tau$ as
$$
	\Psi_\al(\tau,\bx) =\frac{1}{\sqrt{2\pi}|{\cM}|^{1/2}} \int_{\dR\otimes \cM }\D{\bp d\omega} \Psi_\al(\omega,\bp)e^{\ii  \bp \bx }, \qquad
	\al=1,2
$$
The partition function of the considered models has the form
\be
Z = \int D\bar{\Psi}D\Psi
	\,\, e^{S[\Psi,\bar{\Psi}]}\qquad
	S[\Psi,\bar{\Psi}]= \int_{\dR\otimes \cM} \frac{d\omega d^D {\bp}}{2\pi|{\cM}|}\,
		\bar{\Psi}^T(\omega,\bp)\hat{Q}\Psi(\omega,{\bp}),
	\label{Z01}
\ee
%where the products of operators inside expression $Q(p-A(i\partial_p))$ are symmetrized.

As usually, we relate operators $\hat{Q}$ and $\hat{G} = \hat{Q}^{-1}$ defined in Hilbert space ${\cal H}$ of functions (on $\dR\otimes \cM$) with their matrix elements $Q(p,q)$ and ${G}(p,q)$, where the $D+1$ dimensional vectors consist of the spatial parts $\bp, \bq$ and frequencies $p_{D+1}, q_{D+1}$:
$$
Q(p,q) = \langle p|\hat{Q}| q\rangle, \quad {G}(p,q) = \langle p|\hat{Q}^{-1}| q\rangle.
$$
It is implied that the basis of $\cal H$ is normalized as $\langle p| q\rangle = \delta(p_{D+1}-q_{D+1})\delta^{(D)}(\bp-\bq)$. The mentioned operators satisfy
\be
	\hat{Q} \hat{G} = 1
	\label{QG=1}
\ee
or, equivalently,
$$
	\langle p|\hat{Q}\hat{G}|q\rangle = \delta^{(D+1)}({p}-q ).
$$
Eq. (\ref{Z01}) may be rewritten as follows
\be
S[\Psi,\bar{\Psi}]=	
	\int_{\dR\otimes \cM}\! \frac{d^{D+1} {p}_1 }{\sqrt{2\pi|{\cM}|}}
	\int_{\dR\otimes \cM}\!\frac{d^{D+1} {p}_2}{\sqrt{2\pi|{\cM}|}}\,
		\bar{\Psi}^T({p}_1)Q(p_1,p_2)\Psi(p_2)
	\label{Z1}
\ee
while the Green's  function is
\be
{G}_{ab}( k_2, k_1)
= \frac{1}{Z}\int D\bar{\Psi}D\Psi \,
%	 	\frac{\bar{\psi}_b( k_2)}{\sqrt{|{\cM}|}} \frac{\psi_a( k_1)}{\sqrt{|{\cM}|}}
	W_{ab}( k_2, k_1)
	\,\,e^{S[\Psi,\bar{\Psi}]}
	\label{G1}
\ee
where we introduced  the Grassmann-valued Wigner function
\be
	W_{ab}(p,q) = \frac{{\Psi}_b(p)}{\sqrt{2\pi|{\cM}|}}\frac{\bar{\Psi}_a(q)}{\sqrt{2\pi|{\cM}|}}.
	\label{W_{ab}}
\ee
Formally we may also define operator $\hat{W}_{ab}\equiv \hat{W}_{ab}[\Psi,\bar{\Psi}]$, whose matrix elements are equal to the Wigner function,  $W_{ab}(p,q) = \langle p|\hat{W}_{ab}[\Psi,\bar{\Psi}]|q\rangle$.
Indices $a,b$ enumerate the components of the fermionic fields, we will omit them for brevity.

We may consider the $D+1$ dimensional version of Wigner transformation of $\hat G$ in the way similar to that of \Ref{A_W}:
\be
{G}_W(x,p ) \equiv G_W (\tau,\bx;p_{D+1},\bp)
		= \int_\dR dq_{D+1} \int_\cM \D\bq  e^{\ii  (\tau q_{D+1} + \bx  \bq)} {G}({p+q /2}, {p-q /2}).
	\label{GWx}
\ee
Its inverse then is
\be
	{G}(p+q /2,p- q /2) = \frac{1}{2\pi|{\cM}|}\int_\dR d\tau \sum_{\bx\in \cO_{1}} e^{-\ii (\tau \omega + \bx  \bq)} {G}_W(\tau,\bx;\omega,\bp).
	\label{GWxinv}
\ee
In the same way the $D+1$ dimensional Weyl symbol of $\hat Q$ may be defined. For $\hat{Q}  = -\partial_\tau-{\bf H}$ we obtain
$$
{Q}_W(x,p )= \ii\omega-H_W(\bx,\bp)
$$
where $H_W$ is the $D$-dimensional Weyl symbol of the Hamiltonian defined above.

For the slowly varying external electromagnetic field and/or in the presence of weak elastic deformations the function $Q_W(x,p )$ varies slowly as a function of $ \bx $ on the distances of order of the lattice spacing. As a result matrix elements $\Bra{p+\frac q {2}} \hat{Q} \Ket{p-\frac q {2}}$ and $\Bra{p+\frac q {2}} \hat{G} \Ket{p-\frac q {2}}$ are both nonzero in the small vicinity of $\bq=0$. This imposes the bounds on the value of external magnetic field $B$: it should be much smaller than $1/a^2$ (where $a$ is the typical lattice spacing). In practice this means $B \ll 1000$ Tesla. Then we are able to use  Eq. (\ref{ZAB}) and Eq. \Ref{QG=1} becomes a lattice version of the Groenewold equation:
\be
	{G}_W(x,p ) \ast Q_W(x,p ) = 1
	\label{Geq}
\ee
that is
\be
1 =
{G}_W(x,p )
	e^{\frac{\ii}{2} \( \cev{\partial}_{x}\vec{\partial}_p-\cev{\partial}_p\vec{\partial}_{x}\)}
	Q_W(x,p ).
\label{GQW}
\ee
Weyl symbol $Q_W$ of operator $\hat{Q}$ has been calculated above and is given by Eq. (\ref{QW}). For the external fields that vary slowly on the distances of the order of lattice spacing we are able to represent it as a function of $t^{(j)}(x)$ and combination $p-A(x)$:
$$
	Q_W(x,p) = Q_W(t^{(j)}(x),p-A(x)).
$$

\subsection{Expression for the electric current}
\label{Sect4}
Let us consider the variation of the partition function \Ref{Z01} corresponding to the variation of the external field $ A$.

We note first that the action can be written as an operator trace,
\be
	S[\Psi,\bar{\Psi}] = \Tr \(\hat{W}[\Psi,\bar{\Psi}] \hat Q\),
\ee
where $\hat W[\Psi,\bar{\Psi}]$ is the Wigner operator corresponding to \Ref{W_{ab}}.
Vacuum expectation value, defined in the usual way,
\be
	\langle \hat O \rangle = \frac1Z\int D\bar{\Psi}D \Psi \, \hat O e^{S[\Psi,\bar{\Psi}]} ,
\ee
gives then for the variation of the action
\be
\langle \delta S \rangle =
	\int_{\dR\otimes\cM} \!{\frac{dp}{2\pi  |{\cM}|}}
	{\rm tr} \[{G}_W (p,x) \ast \partial_{p_k} Q_W(t^{(j)}(x),p-{  A}(x))	\]\delta{  A}_k(x)
\ee
where we used \Ref{G1} for the expectation value of $\hat W$ and expressed trace of (almost diagonal) operators through a trace of their Weyl symbols
$$
{\Tr} \hat{A} \hat{B} =\Tr(A_W \ast B_W)  = \sum_{x} \int\frac{dp}{(2\pi)^{D+1}}
	\tr (A_W \ast B_W).
$$

Now we obtain
\be
\begin{split}
	\delta {\rm log} \,Z
&=-\int_{\dR^{D+1}}\!\!dx  \int_{\dR\otimes\cM}\! \frac{dp}{2\pi  |{\cM}|} \,
		{\rm tr} \[{G}_W (p,x) \ast \partial_{p_k} Q_W(t^{(j)}(x),p-{  A}(x))
		\]\delta{  A}_k(x).
	\label{dZ2}
\end{split}
\ee
We used that for the slow varying fields
$$
	\sum_{x\in \cO_{1}} \approx \int_{\dR^D} \frac{dx}{|{\cal V}|}
$$
where $|{\cal V}|$ is the volume of the lattice cell. Also we used the following relation between $|{\cal V}|$ and $|{\cM}|$:
$$
|{\cal V}| |{\cM}| = (2\pi)^D.
$$

The total current, i.e. the current density integrated over the whole volume of the system, appears as the response to the variation of $ A$ that does not depend on coordinates:
\begin{eqnarray}
\langle J^k \rangle
%&=&  -\frac{T}{Z}
%	\int D\bar{\Psi}D\Psi \, e^{-S[\Psi,\bar{\Psi}]}
%	\int_{\dR^{D+1}} \D \bx  \, \int_\cM\!{\frac{dp}{(2\pi)^{D+1}}}\,
%		\tr \[{W}_W (p,x) \ast \partial_{p_k} Q_W(p,x)\]
%	\nonumber\\
&=& -T\,
	\int_{\dR^{D+1}} \D x  \, \int_{\dR\otimes\cM}\!{\frac{dp}{(2\pi)^{D+1}}}\,
	\tr \[G_W (p,x) \ast \partial_{p_k} Q_W(p,x)\]
\label{J}
\end{eqnarray}
Here $T$ is temperature that is assumed to be small.	 The properties of the star product allow to rewrite the last equation in the following way:
\be
\langle J^k \rangle = -T\,
	\Tr\[G_W (p,x) \partial_{p_k} Q_W(p-{  A}(\bx))\]
\label{J2}
\ee
This expression for the total current is a topological invariant, i.e. it is not changed when the system is modified continuously. Here $\Tr$ of a Weyl symbol of an operator stands for integration over whole phase space and summation over spinor indices, if any
\be
	\Tr A_W(x,p )\equiv
	\int_{\dR^{D+1}} \D x  \, \int_{\dR\otimes\cM}\!{\frac{dp}{(2\pi)^{D+1}}}\, \tr A_W(x,p ).
	\label{Tr}
\ee

\section{Calculation of the Green's  function in the inhomogeneous lattice models}
\label{GWcalculation}

\subsection{Calculation of the Wigner transformation of the Green's function }

In this section we propose method of calculation of electronic  Green's function in lattice models. This method is based on solving of the Groenewold equation \Ref{GQW}
\be
	Q_W\ast  G_W =1
\label{GE}
\ee
for the Wigner transformation $G_W$ as defined in \Ref{GWx}.  In the following we use Eq. (\ref{QW}) as the definition of the Weyl transform of $\hat Q$. Let us also introduce the following notation
$$
\overleftrightarrow{\Delta} = \frac{i}{2} \left( \overleftarrow{\partial}_{x}\overrightarrow{\partial_p}-\overleftarrow{\partial_p}\overrightarrow{\partial}_{x}\right )
$$
The solution may be written as follows:
\begin{equation}
\begin{aligned}
{G}_W  =&Q^{-1}_W + \sum_{n=1}^{\infty}\sum_{\begin{array}{c} M=1 \\\sum_j k_j= n\\ k_i\ne 0\end{array}}^n \, \frac{(-1)^M}{\Pi_{i=1}^M k_i!}% {k_1! k_2! ... k_M!}
\, \Big[...\Big[Q^{-1}_W \overleftrightarrow{\Delta}^{k_1} Q_W\Big] Q^{-1}_W \overleftrightarrow{\Delta}^{k_2} Q_W\Big]Q^{-1}_W... \overleftrightarrow{\Delta}^{k_M} Q_W \Big] Q^{-1}_W \\
=&\sum_{M=0}^{\infty}  \, \underbrace{\Big[...\Big[Q^{-1}_W(1- e^{\overleftrightarrow{\Delta}}) Q_W\Big] Q^{-1}_W (1-e^{\overleftrightarrow{\Delta}}) Q_W\Big]... (1-e^{\overleftrightarrow{\Delta}}) Q_W \Big]} Q^{-1}_W\\
& \hspace{5cm} \emph{M \,  brackets} \\
=&\sum_{M=0}^{\infty} \,\underbrace{ \Big[...\Big[Q^{-1}_W(1- \ast) Q_W\Big] Q^{-1}_W(1- \ast) Q_W\Big]Q^{-1}_W... (1-\ast) Q_W \Big]} Q^{-1}_W\\
& \hspace{5cm} \emph{M \,  brackets} \\
\label{GQWsff}
\end{aligned}
\end{equation}
In the first row the sum may be extended to the values $M=n=0$, then the first term will be equal to $Q^{-1}_W$.
Let us introduce the product operator $\bullet$, which works as follows being combined with the star product introduced above:
$$
A \bullet B \ast C = (AB) \ast C, \quad A\ast B \bullet C = (A \ast B) \bullet C
$$
In the first equation $\ast$  acts both on $AB$ and on $C$ while in the second equation it acts only on $A$ and $B$. These rules allow to write the above equation in the compact way:
\begin{equation}\begin{aligned}
{G}_W(x,p ) =&\sum_{M=0}^{\infty}  \,\underbrace{Q^{-1}_W(1- \ast) Q_W\bullet Q^{-1}_W(1- \ast) Q_W\bullet Q^{-1}_W... (1-\ast) Q_W \bullet} Q^{-1}_W\\
& \hspace{5cm} {M \, \bullet- products }\\
=&\sum_{M=0}^{\infty}  \,\Big(Q^{-1}_W(1- \ast) Q_W\bullet\Big)^M  Q^{-1}_W
\label{GQWsffb}\end{aligned}\end{equation}
We may  write symbolically:
\revisionZZ{\begin{equation}\begin{aligned}
{G}_W(x,p ) =&\Big(1- \,Q^{-1}_W(1- \ast) Q_W\bullet\Big)^{-1}  Q^{-1}_W = \Big( \,Q^{-1}_W \ast Q_W\bullet\Big)^{-1}  Q^{-1}_W
\label{GQWsffa}\end{aligned}\end{equation}}

In order to show that Eq. (\ref{GQWsff}) is indeed the solution of the Groenewold equation, let us substitute Eq. (\ref{GQWsffa}) to the star product  ${G}_W\ast Q_W $ and obtain
\revisionZZ{
\begin{equation}\begin{aligned}
{G}_W\ast Q_W =& \sum_{M=0}^{\infty} \,\Big(Q^{-1}_W(1- \ast) Q_W\bullet\Big)^M  Q^{-1}_W \ast Q_W \\
& = -\sum_{M=0}^{\infty}\,\Big(Q^{-1}_W(1- \ast) Q_W\bullet\Big)^M  Q^{-1}_W (1-\ast) Q_W
	+ \sum_{M=0}^{\infty}  \,\Big(Q^{-1}_W(1- \ast) Q_W\bullet\Big)^M\\
& =-\sum_{M=0}^{\infty}  \,\Big(Q^{-1}_W(1- \ast) Q_W\bullet\Big)^M
	+ \sum_{M=0}^{\infty}  \,\Big(Q^{-1}_W(1- \ast) Q_W\bullet\Big)^M   \\
& = \Big(Q^{-1}_W(1- \ast) Q_W\bullet\Big)^0 = 1
\label{GQWsffa}
\end{aligned}\end{equation}}

\subsection{Reconstruction of fermion propagator from its Wigner transformation}

Using the definitions of the Wigner transform  \Ref{GWx} and its inverse \Ref{GWxinv}  we find the Green's  function in discrete coordinate space
\be
\begin{split}
{G}(x_1,x_2)
	&=  \frac{1}{{2\pi|{\cM}|}} \int_{\dR\otimes \cM} \D{p _1 d p _2} {G}(p_1,p_2)e^{\ii p_1 x_1-\ii p_2 x _2}\\
	&= \frac{1}{{4\pi^2|{\cM}|^2}}\int d\omega_1 d\omega_2 \int_\cM \D{\bp _1 d\bp _2}   \int d \tau\\
	&\qquad\qquad	\sum_{\bx \in \cO_1}
			e^{-\ii (p_1-p_2) x   + \ii p _1 x _1-\ii p _2 x _2}
			{G}_W \(x, \tfrac{p_1 +p_2}2 \)\\
	&= \frac{1}{2\pi{|{\cM}|}}\int d\omega \int_\cM \D{\bp}  \sum_{\bx \in \cO_{1}} D(\bx-(\bx_1 + \bx _2)/2|\bp) {G}_W (x,p  )
		e^{\ii p   (x_1-x _2)}
\end{split}
\ee
It is assumed that $p_i = (\omega_i,\bp_i)$ and $x = (\tau,\bx)$.
Here
\be
\begin{split}
D(\by|\bp) &
	= \frac{1}{{|{\cM}|}} \int_\cM \D{\bp _1 d\bp _2} \int_{\dR^D}\D \bq e^{-\ii \bq \by  } \delta(\bp-(\bp_1+\bp_2)/2)\delta(\bq-(\bp_1-\bp_2)) \\
& =  \frac{1}{{|{\cM}|}} \int_\cM \D{\bp _1 d\bp _2} e^{-\ii (\bp_1-\bp_2) \by  } \delta(\bp-(\bp_1+\bp_2)/2)
	%\Theta(p_1-p_2\in\cM)
%	\\
%& =  \frac{1}{{|{\cM}|}} \int_\cM \D{\bp _1} e^{-2\ii (\bp_1-\bp) \by  }
%	\Theta(2\bp-\bp_1 \in\cM)
%	\\	
%& =  ? \frac{1}{{|{\cM}|}} \int_\cM \D{\bp _1} e^{-2\iip_1 \by  }
	%\Theta(2p_1\in\cM)\\	
\end{split}
\ee
%$\Theta(P)$  here is the Heaviside function equal to one when  $P$ is true, and zero otherwise.
Notice, that function $D(\by|\bp)$  is not equal to the lattice delta function.

In the particular case, when both hopping parameters and the external electromagnetic field vary slowly, we may substitute the sum over $ \bx  $ by an integral, and $D(\by|\bp) $ by $\delta(\by)$. This gives
\be
{G}(x_1,x_2) \approx
	\frac{1}{2\pi{|{\cM}|}} \int_{\dR\otimes \cM} \D{p}   {G}_W ( (x_1 + x _2)/2, p)e^{\ii p   (x_1-x _2)}
\ee

\section{Total Hall conductance as the topological invariant in phase space}
\label{sigmaTop}
\subsection{Derivation in the framework of Wigner-Weyl formalism}

We discuss here the case when $D=2$ and slightly modify the derivation presented in \cite{ZW2019}. Let us start from Eq. (\ref{J2}) for the electric current. We represent the electromagnetic potential as a sum of the two contributions:
$$
A = A^{(M)} + A^{(E)}
$$
where $A^{(E)}$ is responsible for the electric field while $A^{(M)}$ is responsible for magnetic field. The former is assumed to be weak, and we will keep in Eq. (\ref{J2}) the term linear in $A^{(E)}$.

The Groenewold equation for $G_W$ may be solved iteratively. We will keep in this solution the terms linear in $A^{(E)}$ and in its first derivative. The zeroth order term (that does not contain $A^{(E)}$ at all) is denoted $G_W^{(0)}$. Then
\be
	G_W\approx G_W^{(0)}+  G_W^{(0)}\ast (\partial_{p_m} Q_W A_m) \ast G_W^{(0)}
\ee
Next, we further expand the second term in derivatives of $A$ and write symbolically
\be
	G_W\approx G_W^{(0)}+  G_W^{(1)} + G_W^{(2)}
	\label{GW_dA}
\ee
where
\begin{eqnarray}
	G_W^{(1)} & = &( G_W^{(0)}\ast \partial_{p_m} Q_W^{(0)} \ast G_W^{(0)} ) A^{(E)}_m
	\nonumber \\
	 G_W^{(2)} & = &
 	\frac\ii2 (G_W^{(0)}\ast \partial_{p_m} Q_W \ast  \partial_{p_l} G_W^{(0)})  \partial_{x_l} A^{(E)}_m
	-\frac\ii2 (\partial_{p_l} G_W^{(0)}\ast \partial_{p_m} Q_W^{(0)} \ast   G_W^{(0)})  \partial_{x_l} A^{(E)}_m\nonumber\\
%	& = &-	\frac\ii2 (G_W^{(0)}\ast \partial_{p_m} Q_W^{(0)} \ast G_W^{(0)}\ast \partial_{p_l} Q_W^{(0)} \ast   G_W^{(0)})  \partial_{x_l} A^{(E)}_m
%	+ \frac\ii2 (G_W^{(0)}\ast \partial_{p_l} Q_W^{(0)} \ast   G_W^{(0)}\ast \partial_{p_m} Q_W^{(0)} \ast   G_W^{(0)})  \partial_{x_l} A^{(E)}_m \nonumber\\
	&= &\frac\ii2 (G_W^{(0)}\ast \partial_{p_l} Q_W^{(0)} \ast   G_W^{(0)}\ast \partial_{p_m} Q_W^{(0)} \ast   G_W^{(0)})  F^{(E)}_{lm}
\end{eqnarray}
where we used that $\partial_{p_l} G_W^{(0)} =-G_W^{(0)}\ast \partial_{p_l} Q_W^{(0)} \ast   G_W^{(0)}$.
The $Q_W$ does not depend on the derivatives of $A$, therefore, it is given by
\be
	Q_W =  Q_W^{(0)}+  \partial_{p_m} Q_W^{(0)} A^{(E)}_m
	\label{QW_dA}
\ee

Upon substitution of \Ref{GW_dA} and \Ref{QW_dA} in Eq. (\ref{J2}) the terms proportional to $A^{(E)}$ (i.e. with no derivatives) cancel each other. The remaining term proportional to the field strength $F^{(E)}$ is
\be
\begin{split}
\Tr\, \( G_W^{(2)} \partial_k Q_W^{(0)} \)
	& =\frac{\ii}{2}  {\Tr} \((G_W^{(0)}\ast \partial_{p_l} Q_W^{(0)} \ast   G_W^{(0)}\ast \partial_{p_m} Q_W^{(0)} \ast   G_W^{(0)})  F^{(E)}_{lm}
		\partial_k Q_W^{(0)} \)\\
	&=\frac{\ii}{2}  {\Tr} \((G_W^{(0)}\ast \partial_{p_l} Q_W^{(0)} \ast   G_W^{(0)}\ast \partial_{p_m} Q_W^{(0)} \ast   G_W^{(0)} \ast \partial_k Q_W^{(0)})  F^{(E)}_{lm}
 \)
\end{split}
\ee
We come to the following representation of the average Hall current (i.e. the Hall current integrated over the whole area of the sample divided by this area ${\cal A}$) in the presence of electric field along the $ x_2$ axis:
$$
\langle J_1 \rangle = \frac{\cal N}{2\pi} E_2
$$
Here
\be
{\cal N}
	=  \frac{T \epsilon_{ijk}}{ {\cal A} \,3!\,4\pi^2}\,  \Tr
		\[
		{G}_W(x,p )\ast \frac{\partial {Q}_W(x,p )}{\partial p_i} \ast \frac{\partial  {G}_W(x,p )}{\partial p_j} \ast \frac{\partial  {Q}_W(x,p )}{\partial p_k}
		\]_{A^{(E)}=0}
  \label{calM2d230}
\ee
with $\Tr$ defined in \Ref{Tr}. This expression for $\cal N$ is a topological invariant in phase space, i.e. it is not changed if the system is modified smoothly within a finite region distant from the boundary of the sample or from infinity if the sample is infinite. This may be checked via the direct consideration of a variation of Eq. (\ref{calM2d230}) with respect to the variation of $Q_W$.

\subsection{From topological invariant in phase space expressed through $G_W, Q_W$ to the standard expression for Hall conductance}

In the previous section we showed that Hall conductance (i.e. the conductivity integrated over the whole area of the sample) is given by $\sigma_{xy} = {\cal N}/2 \pi$ where ${\cal N}$ is the topological invariant in phase space.
Our derivation is applicable to the general case of the inhomogeneous one-particle Hamiltonian including the case when elastic deformations are present.  Our next purpose is to bring Eq. (\ref{calM2d230}) to the conventional expression for the Hall conductance in the case, when the non-interacting charged fermions have Hamiltonian ${\cal H}$.

First of all, one may show that Eq. (\ref{calM2d230}) is equivalent to the following representation for $\cal N$ in terms of the Green's 's function written in momentum representation:
\begin{eqnarray}
{\cal N} &=& \frac{T \,(2\pi)^3}{{\cal A}\, 3!\,4\pi^2} \, 	\epsilon_{ijk} \int  \prod_{l=1}^4 d^3p^{(l)}
	\tr \[
	{G}(p^{(1)},p^{(2)})\Big( [\partial_{p^{(2)}_i} + \partial_{p^{(3)}_i}] Q(p^{(2)},p^{(3)})\Big)	\right. \nonumber\\
	&& 	
	\qquad  \times \left. \Big( [\partial_{p^{(3)}_j} + \partial_{p^{(4)}_j}]  G(p^{(3)},p^{(4)}) \Big)
		\Big( [\partial_{p^{(4)}_k} + \partial_{p^{(1)}_k}] Q(p^{(4)},p^{(1)})\Big)
	\]_{A=0}
	\label{calM2d23P}
\end{eqnarray}
This may be proved noticing that the functional trace of a product of two operators is expressed through their Weyl symbols as follows:
$$
{\Tr} \hat{A} \hat{B} =\Tr(A_W \ast B_W)  = \int d^3 x  \int \frac{d^3 p}{(2\pi)^3}
	\tr (A_W \ast B_W)
$$
(Again, we need that the matrix elements $\Bra{p+\frac{q} {2}} \hat{A} \Ket{p-\frac{q} {2}}$ and $\Bra{p+\frac{q} {2}} \hat{B} \Ket{p-\frac{q} {2}}$ are nonzero only when $q$ remains in the small vicinity of zero.) Applying this formula several times to Eq. (\ref{calM2d230}) we come to Eq. (\ref{calM2d23P}).

Secondly, for non-interacting fermions described by ${\cal H}$ with energy eigenstates $|n\rangle$: $\cH |n\rangle = \cE_n |n\rangle$,  function $Q(p^{(1)},p^{(2)})$ in momentum space has the following form:
\be
Q(p^{(1)},p^{(2)}) \equiv \langle p^{(1)}| \hat{Q} | p^{(2)}\rangle
	= \( \delta^{(2)} (\bp^{(1)}-\bp^{(2)}) \ii \omega^{(1)}
		- \langle \bp^{(1)}| {\cal H} | \bp^{(2)}\rangle \) \delta(\omega^{(1)}-\omega^{(2)})
\ee
where $p = (p_1,p_2,p_3) = (\bp ,\omega)$. At the same time
$$
G(p^{(1)},p^{(2)}) = \sum_{n} \frac{1}{\ii\omega^{(1)}-{\cal E}_n} \langle \bp ^{(1)}| n \rangle \langle n | \bp ^{(2)}\rangle \delta(\omega^{(1)}-\omega^{(2)})
$$
This way we obtain:
\rev{\begin{eqnarray}
{\cal N} = -\frac{\ii\,(2\pi)^2}{8\pi^2 \, {\cal A}}\,\sum_{n,k} \int_\dR&& \!d \omega  \prod_{l=1}^4 d^2\bp ^{(l)}
 \epsilon_{ij}
  \nonumber\\
 &&
 \tr\Big[
 	\frac{1}{(\ii\omega^{}-{\cal E}_n)^2} \langle \bp ^{(1)}| n \rangle \langle n | \bp ^{(2)}\rangle
 		\Big( [\partial_{p^{(2)}_i} + \partial_{p^{(3)}_i}] \langle \bp ^{(2)}| {\cal H} | \bp ^{(3)}\rangle\Big)
 		\nonumber\\&&
 		\qquad \frac{1}{(\ii\omega^{}-{\cal E}_k)} \langle \bp ^{(3)}| k \rangle \langle k | \bp ^{(4)}\rangle
 		\Big( [\partial_{p^{(4)}_j} + \partial_{p^{(1)}_j}]  \langle \bp ^{(4)}| {\cal H} | \bp ^{(1)}\rangle \Big)
 \Big]_{A=0}\nonumber
\end{eqnarray}}
One may represent
$$
	[\partial_{p^{(4)}_j} + \partial_{p^{(1)}_j}]  \langle \bp ^{(4)}| {\cal H} | \bp ^{(1)}\rangle% =\langle  p ^{(4)} | \frac{\partial }{\partial {\hat{\bp}^{i}}}{\cal H} | p ^{(1)} \rangle
= i   \langle \bp ^{(4)}| {\cal H} {\hat \bx }_j -{\hat \bx }_j{\cal H}   | \bp ^{(1)}\rangle
 = i \langle \bp ^{(4)}| [{\cal H}, {\hat \bx }_j]| \bp ^{(1)}\rangle.
$$
By operator $\hat \bx$ we understand $i\partial_{\bp}$ acting on the wavefunction written in momentum representation:
$$
\hat{x}_j \Psi(\bp)
	= \langle \bp|\hat{x}_j |\Psi\rangle
	= i\partial_{p_j} \langle \bp|\Psi\rangle
	= \ii \partial_{p_j} \Psi(\bp)
$$
Then, for example,
$$
\hat{x}_j \delta^{(2)}(\bq-\bp)
	= \langle \bp|\hat{x}_j |\bq\rangle
	= \ii\partial_{p_j} \langle \bp|\bq \rangle
	= \ii \partial_{p_j} \delta^{(2)}(\bp-\bq)
	= -\ii \partial_{p_j}\langle \bq|\bp \rangle
$$
Therefore, we can write
$$
\hat{x}_j |\bp\rangle = -\ii\partial_{p_j} |\bp \rangle
$$
Notice, that the sign minus here is counter-intuitive because the operator $\hat \bx $ is typically associated with $+\ii\partial_\bp$. We should remember, however, that with this latter representation the derivative acts on $\bp$ in the bra-vector $\langle \bp| $ rather than on $\bp$ in $| \bp \rangle$. Above we have shown that the sign is changed when the derivative is transmitted to $\bp$ of $| \bp \rangle$.

Thus we have
\begin{eqnarray}
{\cal N} &=&  \frac{\ii\,(2\pi)^2}{8\pi^2\, {\cal A}}\,\sum_{n,k} \int_\dR \D\omega  \epsilon_{ij}\,
		\frac{\langle n| [{\cal H}, {\hat \bx }_i] | k \rangle  \langle k | [{\cal H}, {\hat \bx }_j] | n \rangle  }
			{(\ii\omega^{}-{\cal E}_n)^2 (\ii\omega^{}-{\cal E}_k)}
	\nonumber\\
&=&-\frac{2\ii\,(2\pi)^3}{8\pi^2\, {\cal A}}\,\sum_{n,k}   \, \epsilon_{ij}\,
	\frac{\theta(-{\cal E}_n)\theta({\cal E}_k)}{({\cal E}_k-{\cal E}_n)^2}
		  \langle n| [{\cal H}, {\hat \bx }_i] | k \rangle    \langle k | [{\cal H}, {\hat \bx }_j] | n \rangle  .
	\label{sigmaHH}
\end{eqnarray}
The last expression is just the conventional expression for the Hall conductance (\rev{multiplied by $2\pi$}) for the given system. Notice, that it is valid for the slowly varying electromagnetic potential only (the potential almost does not vary at the distance of order of lattice spacing). Then operator ${\hat \bx} = i\frac{\partial}{\partial \bp}$ has the meaning of coordinate operator.

\section{Integer Quantum Hall effect in the presence of varying magnetic field and elastic deformations}
\label{IQHE}

%\section{Hall conductance in graphene in the presence of magnetic field and weak variations of hopping parameters}
%\label{GrapheneHall}
\subsection{Constant magnetic field and constant hopping parameters}

In this subsection we repeat the standard derivation of the Hall conductance in the noninteracting $2D$ models with constant magnetic field perpendicular to the surface, and constant hopping parameters. It is assumed here that the magnetic field $\bB$ is sufficiently weak, so that $|\bB| a^2 \ll 1$, where $a$ is the lattice spacing. Then
the Hall conductivity may be represented as ${\cal N}/(2\pi)$, where $\cal N$ is given above in \Ref{sigmaHH}.

In order to calculate the value of $\cal N$ we decompose the coordinates $ \bx _1, \bx _2$
in relative coordinates $\xi_i$ (with bounded values) and center coordinates $X_i$ (the unbounded part)
\be
\hat{x}_1 = \hat{\xi}_1 + \hat{X}_1,\qquad
	 \hat{x}_2 =   \hat{\xi}_2 + \hat{X}_2
	 \label{xi-X}
\ee
where
\be
\hat{\xi}_1 = -\frac{\hat{ p}_2-B  x _1}{B},\qquad \hat{X}_1 = \frac{\hat{ p}_2}{ B}
\ee
\be
\hat{\xi}_2 = -\frac{\hat{ p}_1}{ B},\qquad \hat{X}_2 = \frac{\hat{p}_1- B  x _2}{ B}
\ee
Then the commutation relations follow:
\be
[\hat{\xi}_1,\hat{\xi}_2] = \frac{i}{ B}, \quad [\hat{X}_1,\hat{X}_2] =-\frac{i}{ B},
\qquad
	[\hat{\xi}_i,\hat{X}_j]=0\quad \forall i,j
	\label{comm-xi-X}
\ee
%$$
%[{\cal H}, \xi_1] = -i  \frac{\partial}{\partial p_x} {\cal H}, \quad [{\cal H}, \xi_2] =  -i \frac{\partial}{\partial %p_y} {\cal H}
%$$
Since the Hamiltonian is a function of $\xi_i$ only (in Landau gauge)
\be
	{\cal H}\equiv \cH(\xi_1,\xi_2)
\ee
its commutator with $X_j$ vanishes
\be
[{\cal H}, \hat{X}_1] =  [{\cal H}, \hat{X}_2] =  0.
\ee
We use these relations to obtain:
\be
\begin{split}
{\cal N} &=  -\frac{2\ii\,(2\pi)^3}{8\pi^2\, {\cal A}}\,\sum_{n,k}
	\Big[
		\frac{1}{({\cal E}_k-{\cal E}_n)^2}  \langle n| [{\cal H}, {\hat \xi}_i] | k \rangle
		\langle k | 	[{\cal H}, {\hat \xi}_j] | n \rangle
	\Big]_{A=0}
\epsilon_{ij} \, \theta(-{\cal E}_n)\theta({\cal E}_k) \\
&=   \frac{2\ii\pi}{{\cal A}}\,\sum_{n,k}  \epsilon_{ij}\,
	\Big[
		\langle n|  {\hat \xi}_i | k \rangle    \langle k |  {\hat \xi}_j | n \rangle
	\Big]_{A=0}\theta(-{\cal E}_n)\theta({\cal E}_k)
 \\
&=   \frac{2\ii\pi}{{\cal A}}\,\sum_{n}   \,\Big[  \langle n|  [{\hat \xi}_1,  {\hat \xi}_2 ]| n \rangle  \Big]_{A=0}\theta(-{\cal E}_n)
 \\
&=   -\frac{2\ii\pi}{{\cal A}\, B}\,\sum_{n}   \,   \langle n|   n \rangle  \theta(-{\cal E}_n).
\end{split}
\label{N-nn}
\ee

Momentum $p_2$ is a good quantum number, and it enumerates the eigenstates of the Hamiltonian:
$$
{\cal H} |n\rangle = {\cal H}(\hat{ p}_1, \hat p_2-{  B} \hat x _1)|p_2, m\rangle = {\cal E}_{m}(p_2)|p_2, m\rangle , \, m\in Z
$$
We assume that the size of the system is $L\times L$.
This gives
\begin{eqnarray}
{\cal N} &=&-\frac{(2\pi)}{{\cal A}}\,\sum_{m}\int \frac{dp_2 L}{2\pi}  \, \frac{1}{  B} \theta(-{\cal E}_m(p_2))
  \label{calM2d232}
\end{eqnarray}
Average value $\langle \bx  \rangle = p_y/{  B}$  plays the role of the center of orbit, and this center should belong to the interval $(-L/2, L/2)$. This gives
\begin{eqnarray}
{\cal N}
	&=&   N  \, {\rm sign}(-{  B}),
	\label{calM2d233}
\end{eqnarray}
 Here ${\cal A} = L^2$ is the area of the system while $N$ is the number of the occupied branches of spectrum. This way we came to the conventional expression for the Hall conductance of the fermionic system in the presence of constant magnetic field and constant electric field.

It is worth mentioning, that for graphene in addition to the considered above expression of $\sigma_{xy}$ there is another contribution. It is caused by the deep energy levels, which are not described by the presented here theory. According to \cite{Hatsugai2} this contribution may be calculated when the tight-binding model of graphene is considered exactly. This is possible for constant magnetic field in the absence of elastic deformations. It appears that this contribution of the  deep levels cancels precisely that of ${\cal N}/(2\pi)$ at the half filling. We denote this term $\sigma_{xy}^{(0)} = {\cal N}^{(0)}/(2\pi)$, and the final expression for the Hall conductivity becomes
\begin{equation}
\sigma_{xy} = \frac{{\cal N}}{2\pi}-\sigma_{xy}^{(0)}\label{sigmadeformed}
\end{equation}
Correspondingly, in Eq. (\ref{calM2d233}) $N$ is counted from a certain deep Landau Level in such a way, that we have
\begin{equation}
\sigma_{xy} = \frac{N^\prime}{2\pi} {\rm sign}(-{B})\label{sigmadeformed2}
\end{equation}
where $N^\prime$ is counted from the half filling (the LLL being occupied contributes with the factor $1/2$).

\subsection{Constant magnetic field and weakly varying hopping parameters}
\label{Secttt}
Let us consider the case when
$$
	t^{(j)}(\bx) = t^{(j)}_0 e^{ \bx  \bbf }.
$$
(with some constant spatial vector $ \bbf$) for specific Hamiltonian, which we define by its Weyl symbol
\begin{equation}
{\cal H}_W(\bx,\bp ) = \sum_{j=1}^M \left(\begin{array}{cc}-\mu/N & t^{(j)}(\bx)\, e^{\ii  (p  \bbj-{ A}^{(j)}(x ))} \\
 t^{(j)}(\bx) \, e^{-\ii (p \bbj-{ A}^{(j)}(\bx))} & -\mu/N \end{array}\right)\label{HWm}
\end{equation}
here $\mu$ is chemical potential. As above, we decompose the coordinates $ \bx _1, \bx _2$ into relative coordinates $\xi_i$ and the center coordinates $X_i$ using \Ref{xi-X}.  We still have the commutation relations \Ref{comm-xi-X}. However, since $t^{(j)}(\bx)$ now depend on coordinates, the Hamiltonian does not commute anymore with $X_1$, $X_2$. Instead we have
\be
	{\cal H}_W \ast {X}_{W,j}-{X}_{W,j}\ast {\cal H}_W  = \frac{i}{2 B} \epsilon^{ji}f_i \, ({\cal H}_W-\mu) .
\ee
Here
\be
 \hat{X}_1 = \frac{\hat{p}_2}{ B},\qquad \hat{X}_2 = \frac{\hat{p}_1- B \hat x _2}{ B}
\ee
and $X_{W,i}$ is their Weyl symbols.
This gives
\be
	{\cal H} \hat{X}_{j}-\hat{X}_{j} {\cal H}  = \frac{\ii}{2 B} \epsilon^{ji}f_i \, ({\cal H}-\mu)
\ee
Then for $n\ne k$ we obtain:
$$
	\langle n | {\cal H} \hat{X}_{j}-\hat{X}_{j} {\cal H}|k \rangle  = 0
$$
and we come again to \Ref{N-nn}
\be
{\cal N} =   -\frac{2\pi}{{\cal A}\, B}\,\sum_{n}   \,   \langle n|   n \rangle  \theta(-{\cal E}_n)
	=   -\frac{2\pi}{ B } \,{\rho}
\ee
where now 	$\rho$ is the average density of occupied states.

If we require, in addition, that $t^{(j)}$ does not depend on $y$ and depends on $ \bx $ only, then momentum $p_y$ is still a good quantum number, and \Ref{calM2d232} is applicable. As above, we will obtain
\begin{eqnarray}
{\cal N}
&=&  N  \, {\rm sign}(-{B}),
  \label{calM2d233}
\end{eqnarray}
Recall that  ${\cal A} = L^2$ is the area of the system while $N$ is the number of the occupied branches of spectrum. One can see, that in the presence of constant magnetic field and the hopping parameters that depend on $ \bx $ and do not depend on $y$ (i.e. $t^{(j)}(\bx) = t^{(j)}_0 e^{ x_1  f_1}$) the Hall conductivity is given by the same standard value as for the constant hopping parameters. Here we assume also, that weak elastic deformations are not able to modify the contribution $\sigma^{(0)}_{xy}$ of the deep Landau Levels, so that Eq. (\ref{sigmadeformed}) remains valid, and we come finally to the  conductivity   (averaged over the area of the sample) is
\begin{eqnarray}
\sigma_{xy} &=&  \frac{ N^\prime}{2\pi}  \, {\rm sign}(-{B}),
  \label{calM2d233}
\end{eqnarray}
where the number of occupied branches of spectrum $N^\prime$ is counted from the half-filling.

\subsection{Weak variations of magnetic field and hopping parameters}

\label{SectWD}

The key point of our calculation is that the total integrated Hall conductance is given by the topological invariant in phase space $\sigma_{xy} = {\cal N}/2 \pi$ where ${\cal N}$ is given by \Ref{calM2d230}
$$
{\cal N}
=  \frac{\epsilon_{ijk}}{{\cal A}\, 3!\,4\pi^2}\,  \Tr
\[
{G}_W(\bx,\bp )\ast \frac{\partial {Q}_W(\bx,\bp )}{\partial p_i} \ast \frac{\partial  {G}_W(\bx,\bp )}{\partial p_j} \ast \frac{\partial  {Q}_W(\bx,\bp )}{\partial p_k}
\]_{A =0}
$$
From the above consideration, we also know that for the constant magnetic field
\begin{eqnarray}
{\cal N}
&=&  N  \, {\rm sign}(-{B})\label{NBc}
\end{eqnarray}
Smooth modification of Hamiltonian does not change the value of $\cal N$ until the phase transition is encountered. We need, however, that both hopping parameters and the electromagnetic field remain equal to their original values at spatial infinity $|x| \to \infty$. We come to the conclusion, that under this condition the total Hall conductance is still given by Eq. (\ref{NBc}) for the weakly varying hopping parameters and magnetic field.

\subsection{Analytical elastic deformations in graphene}

In graphene the relation between $t$, $A$  and $u$ is given by \cite{VZ2013},
\be
\begin{split}
	t^{(j)}(\bx)=t[1-{\beta} u_{kl}({\bx})  \bj_k \bj_l],\qquad
{A}_1  =-\frac{\beta}{a}\,u_{12}\quad
{A}_2  =  \frac{\beta}{2a}\,(u_{22}-u_{11})
\end{split}
\label{t-A-u}
\ee
With elementary translations given by \Ref{l-s-gr}, the nontrivial part of $\tj$ is
\be
u_{kl}({\bx})  \bj_k \bj_l =
	\frac{a^2}4 \(\begin{array}{c}
		4 u_{11}\\
		u_{11}+2\sqrt3 u_{12}+3 u_{22}\\
		u_{11}-2\sqrt3 u_{12}+3 u_{22}
	\end{array}\)
\ee
Requiring that
\be
	t^{(1)}(\bx) = t^{(2)}(\bx) = t^{(3)}(\bx),
	\label{t-cond}
\ee
 consistent with Sect. \ref{Secttt}, we come to the Cauchy-Riemann conditions
$$
	\frac{\partial u_1}{\partial x_1} = \frac{\partial u_2}{\partial x_2}, \qquad
	\frac{\partial u_2}{\partial x_1}  =-\frac{\partial u_1}{\partial x_2}
$$
that is $h(z) \equiv u_1(z) + \ii u_2(z)$ is analytic as a function of $z = x_1 +\ii x_2 $.
\footnote{
There is another solution of \Ref{t-cond}
$$
\partial_1 u_1 = -2-\partial_2 u_2, \quad \partial_1 u_2  = \partial_2 u_1,
$$
which however brakes the smallness condition $\beta |u_{ij}|\ll 1$.}

In this case we have vanishing emergent gauge field, while up to the terms linear in the derivatives of the hopping parameters the results of Sect. \ref{Secttt} give
\begin{equation}
{\cal N}  = -\frac{2\pi}{ { B}} \,  {\rho}\label{Nrho}
\end{equation}
where $\rho$ is the average density of occupied states. On the other hand, the results of Sect. \ref{SectWD} ensure, that  any weak variations of both hopping parameters and magnetic field give
$$
{\cal N}  =-N\,{\rm sign}\,{  B}
$$
where $N$ is the number of occupied Landau levels (now instead of the degenerate Landau level we may have the energy band parametrized by certain parameters).
Comparing this result with  Eq. (\ref{Nrho}) we obtain
$$
  \rho = \frac{|{  B}|}{2\pi} N
$$
for the elastic deformations given by analytical function of coordinates (i.e. when the emergent magnetic field is absent).

%\appendix

\section{Conclusions and discussions}

\label{SectConcl}

In the present paper we proceed with the development of Wigner-Weyl formalism for the tight-binding models of solid state physics  (or, equivalently, for the lattice regularized quantum field theory). We extend the previous works made in this direction \cite{ZW1,ZW2,ZW3,ZW4,ZW5,ZW6,ZW2019,ZZ2019,SZ2019}. The developed technique is applied to the class of the inhomogeneous models that includes, in particular, the tight-binding model of graphene in the presence of both inhomogeneous magnetic field and nontrivial elastic deformations. It is worth mentioning, that majority of our results may be applied to other models of solid state physics. Apart from the family of two-dimensional honeycomb lattice materials (graphene, germanene, silicene, etc), all rectangular lattice crystals, both in two and three dimensions can be treated with developed methods, if described within nearest-neighbour approximation.  In these cases the electrons may jump only between the nearest neighbors  and there is the $Z_2$ sublattice symmetry. The lattice consists of the two sublattices $\cO_1$ and $\cO_2$. For each $ \bx  \in \cO_1$ site $ \bx  + \bbj  \in \cO_2$ with fixed vectors $\bbj $, where $j = 1,2,...,M$. For the honeycomb lattice $M = 3$, for the $2D$ rectangular lattice $M = 4$, for the $3D$ rectangular lattice $M = 8$. Among the mentioned models only the two-dimensional model on the honeycomb lattice describes sufficiently accurately the real system (graphene). Therefore, the emphasis is on the application to the physics of graphene. The particular interest in our study is the consideration of arbitrarily varying external magnetic field and nonhomogeneous elastic deformations.

We obtain the following main results:
\begin{enumerate}

\item{} We calculate Weyl symbol of lattice Dirac operator (i.e. the operator $\hat Q$ that enters the action $\sum_{\bx,\by } \bar{\Psi}_x Q_{\bx,\by } \Psi_y$) in the presence of both elastic deformations and slowly varying external electromagnetic field:
\be
{Q}_W =
i \omega-t  \sum_j
	\( 1- \beta u_{kl}({\bx})  \bj_k \bj_l \)
	\(\begin{array}{cc}
		0 & e^{\ii  (\bp  \bbj -{ A}^{(j)}({\bf r} ))} \\
		e^{-i (\bp  \bbj -{ A}^{(j)}({\bf r} ))} & 0
	\end{array}	\)\label{QWc}
\ee
where
 $u_{ij}$ is the tensor of elastic deformations while
 $$
	A^{(j)}(\bx) = \int_{\bx-\bbj /2}^{\bx + \bbj /2}  \bA(\by)d\by.
$$
It is assumed that the variation of electromagnetic field $A(\bx)$ at the distances of order of the lattice spacing may be neglected. In practise this corresponds to magnetic fields ${B}$ that obey ${B}a^2 \ll 1$.
In practice this bound reads ${B} \ll 1000$ Tesla. Also we require that the typical wavelenth of the external electromagnetic field is much larger than the lattice spacing. This does not allow to use Eq. (\ref{QWc}) for matter interacting with the X-rays with the wavelengths of the order of several Angstroms and smaller.

\item{} Wigner transformation of electron propagator in the presence of slowly varying magnetic field and arbitrary elastic deformations may be calculated using the following expression:
\revisionZZ{\begin{equation}\begin{aligned}
{G}_W(x,p )
=&\sum_{M=0...\infty}  \, \underbrace{\Big[...\Big[Q^{-1}_W(1- e^{\overleftrightarrow{\Delta}}) Q_W\Big] Q^{-1}_W (1-e^{\overleftrightarrow{\Delta}}) Q_W\Big]... (1-e^{\overleftrightarrow{\Delta}}) Q_W \Big]} Q^{-1}_W\\
& \hspace{5cm} \emph{M \,  brackets} \label{GWc}
\end{aligned}\end{equation}}
where
$$
\overleftrightarrow{\Delta} = \frac{i}{2} \left( \overleftarrow{\partial}_{x}\overrightarrow{\partial_p}-\overleftarrow{\partial_p}\overrightarrow{\partial}_{x}\right )
$$

\item{} Electron propagator in the presence of slowly varying electromagnetic field and elastic deformations may be expressed through the Wigner transformed Green's  function as follows:
\begin{eqnarray}
{G}(x_1,x_2) &\approx&
\frac{1}{2\pi{|{\cM}|} }\int d p   {G}_W ( (x_1 + \bx _2)/2, p)e^{-i p  (x_1-\bx _2)}
\end{eqnarray}
 where $|\cM|$ is the area of the Brillouin zone.

\item{} We prove that the total average Hall conductivity (i.e. the Hall conductivity integrated over the area of the sample and divided by this area, in the presence of varied weak magnetic field ${\cal B} \ll 1/a^2$ and elastic deformations)  has the form of $\sigma_{xy} = \frac{\cal N}{2 \pi}-\frac{{\cal N}^{(0)}}{2\pi}$ where ${\cal N}$ is the topological invariant in phase space
\be
{\cal N} =  \frac{T}{{\cal A}\,3!\,4\pi^2}\,  \epsilon_{ijk} \,{\Tr}
	\[  {G}_W(p,x)\ast \frac{\partial {Q}_W(p,x)}{\partial p_i} \ast \frac{\partial  {G}_W( p,x)}{\partial p_j} \ast \frac{\partial  {Q}_W(p,x)}{\partial p_k}
	\]
	\label{calM2d23c}
\ee
where $\cal A$ is the area of the sample while in graphene ${\cal N}^{(0)}$ is the value of ${\cal N}$ at half-filling (with constant magnetic field and without elastic deformations).
Thus we extend here the results of \cite{ZW2019} to the case, when in addition to the inhomogeneous magnetic field an arbitrary elastic deformation is present.
The resulting expression works for the magnetic field slowly varying in the limited region of the sample, such that it approaches constant value $B$ close to the boundary of the sample.

\item{} The above mentioned representation of the average Hall conductivity through the topological invariant in phase space allows to prove that in graphene it is robust to both weak variations of magnetic field and weak elastic deformations. It is worth mentioning that both mentioned variations of magnetic field and elastic deformations are to be concentrated within the finite region far from the boundary of the sample. Under these conditions  Eq. (\ref{calM2d23c}) is not changed for the smooth variations of lattice Hamiltonian (for the proof see Appendix D in \cite{ZW2019}).

\item{} The special case of elastic deformations is considered, when the emergent gauge field in graphene is absent. It is shown that the corresponding deformations are given by the arbitrary analytical functions of coordinates. Namely, the condition of the absence of emergent gauge field is equivalent to the Riemann-Cauchy conditions for the displacement function $u_i$, $ i= 1,2$. As a result the function $u(z) = u_1(z) + i u_2(z)$ appears to be analytical function of $z=x_1 + i \bx _2$, where $ \bx _i$ are the coordinates of the carbon atoms in the unperturbed honeycomb lattice.  Under these circumstances for the constant magnetic field $B$ the Hall current is given by
 \begin{equation}
	 I_{xy} =- \frac{N^\prime \, U}{2\pi}\,{\rm sign}\,{B}
	 \label{Ixy}
\end{equation}
where $U$ is voltage while $N^\prime$ is the number of occupied Landau levels (counted from the half filling). Now unlike the case of the unperturbed graphene the Landau levels may already not be degenerate.

\end{enumerate}

It is worth mentioning that our results were obtained in the absence of both disorder and Coulomb interactions. According to the standard considerations (see also Section 6 of \cite{ZW2019}) the total Hall current is typically robust to the introduction of disorder. Let us discuss in this respect  the case of graphene in the presence of constant magnetic field but without strain. In pure ideal graphene without impurities all Landau levels participate in the QHE. However, the theory developed here is valid for the levels sufficiently close to the half-filling. The consideration of the model in the absence of elastic deformations and for the constant magnetic field demonstrated, that deep energy levels contribute to the total conductance in a very peculiar way (see \cite{Hatsugai2} and references therein). These contributions contain several jumps, and make the conductance negative for the Fermi energy placed somewhere below zero. Starting from a certain level below zero our theory works, and it gives contributions to conductivity proportional to the number of occupied levels (see, for example, calculation in Sect. \ref{IQHE} above). The sum of this contribution and the contribution of deep energy levels results in vanishing conductance at the half filling. As a result the Landau Levels (LL) participate in the QHE being counted from the neutrality point. The occupied levels above the neutrality point represent the so-called particle LL's, while the vacant ones below neutrality point represent the so-called hole LL's. Then, in  Eq. (\ref{Ixy}) the value of $N$ is negative for the hole LL's and positive for the particle LL's.

In the presence of disorder the Hall current density is pushed towards the boundary.
It appears that the neutrality point (when chemical potential is in the middle of the Lowest Landau Level) corresponds to vanishing Hall conductivity. Again, the Landau Levels (LL) participate in the QHE being counted from the neutrality point.  This occurs now because close to the boundary the branches of energy spectrum above and below the neutrality point behave differently. Energies of those above the neutrality point are increased while energies of the branches situated below it are decreased. As a result there is no crossing of the energy levels with the Fermi level on the boundary at neutrality point \cite{Tong:2016kpv}, and, consequently, there are no gapless edge states that are to be the carriers of the Hall current. This shows, that  weak disorder does not cause a jump in the value of total conductance. The average conductivity  (the conductivity integrated over the area of the sample divided by this area) is given by $\sigma_{xy} = \frac{ N^\prime}{2\pi}\,{\rm sign}\,{B}$, where $N^\prime$ is the number of {\it occupied electronic energy levels counted from the half filling}. Therefore, for graphene $N^\prime$ may be both negative and positive. Moreover, for the chemical potential just above zero only half of the Lowest Landau Level contributes the Hall conductance. Therefore, in this case $N^\prime = \frac{1}{2} \, g_s g_v = \frac{4}{2} = 2$.  Next, we may turn on weak variations of magnetic field and weak elastic deformations. The value of the average Hall conductivity should remain the same until the topological phase transition to the state with a different value of Hall conductivity is encountered. For sufficiently strong elastic deformations and/or variations of magnetic field the very notion of Landau levels may loose its sense, but the values of Hall conductivity may still remain nonzero.

We expect that the Hall current is robust to the weak Coulomb interactions (at least in the presence of a sufficient amount of disorder) although the detailed investigation of this issue is still to be performed (see \cite{ZZ2019} and references therein), especially in the presence of elastic deformations and variations of magnetic field. At the same time, the clean samples of graphene (very weak disorder) exhibit the fractional quantum Hall effect (FQHE) due to the Coulomb interactions. The investigation of this issue remains out of the scope of the present paper although we expect that Eq. (\ref{calM2d23c}) may still be related  somehow to the description of the FQHE.

We suppose that the results obtained here may be used further in the investigation of various properties of graphene. In particular, Eqs. (\ref{QWc}), (\ref{GWc}) determine electron propagator in the complete tight-binding model in the presence of both elastic deformations and slowly varying external electromagnetic field. This propagator may be used in those investigations of transport properties that require use of the complete tight-binding model, i.e. when the low energy effective continuum field theory of graphene is not sufficient for the solution of a particular problem. Since the form of the obtained expressions is rather complicated, and the result of Eq. (\ref{GWc}) is represented in the form of the infinite series, the practical applications of the obtained formulas are likely to require certain numerical techniques.

%Since the Coulomb interactions are sufficiently strong (in the suspended graphene the effective fine structure constant is of the order of unity), we also expect, that the obtained result for the electron propagator may become an important ingredient of Hybrid Monte-Carlo algorithm to be used for the numerical lattice simulations.

We also expect that the practical calculation of Hall conductivity using Eq. (\ref{calM2d23c}) may require the application of certain numerical procedures. The possible problem to be solved using this expression is the calculation of Hall conductivity in the presence of varying magnetic field and/or varying elastic deformations. For the constant external magnetic field and without elastic deformations the result for the Hall conductance is well-known. According to our results weak elastic deformations and weak variations of magnetic field cannot affect the value of the total Hall current. However, when the variations become stronger, the system may undergo a topological phase transition to the state with different value of Hall conductance. We may determine the critical values of magnetic field variation and/or deformation tensor variation using the direct evaluation of an integral in Eq. (\ref{calM2d23c}). Both numerical and analytical methods of this evaluation await for their development.

It is worth mentioning, that the simplified version of Eq. (\ref{calM2d23c}) (discussed in \cite{ZW1}) that appears when $G_W$ does not depend on coordinates, represents the generator of the co-homology group ${\cal H}^{(3)}({\cM})$, where ${\cM}$ is momentum space. Eq. (\ref{calM2d23c}) also awaits for the interpretation using the language of algebraic topology. At the present moment we notice only that this topological invariant certainly plays a role in the classification of the homotopic classes of maps $G: {\cM}\otimes {\cal R} \rightarrow GL(2,C)$, where ${\cal R}$ is the coordinate space with certain boundary conditions while $\cM$ is momentum space.

We would like to notice again, that the theory presented here is valid for the slowly varying potentials, which is consistent with the requirement ${B} a^2 \ll 1$. It would be interesting to extend the Wigner-Weyl formalism to the precise consideration of the tight-binding model of graphene in the presence of strong magnetic fields ${B} \sim 1/a^2$. Another challenge is an extension of our results to the investigation of the fractional Hall effect.

The authors are grateful for sharing ideas, comments and collaboration in the adjacent fields to M.Suleymanov, Xi Wu, and Chunxu Zhang. M.A.Z. is indebted for valuable discussions to G.E.Volovik.

\bibliographystyle{unsrt}
\bibliography{wigner2}

\end{document}